\documentclass[twocolumn,showpacs,amsmath,amssymb]{revtex4}

\usepackage{graphicx}
\usepackage{dcolumn}
\usepackage{bm}
\usepackage{xcolor}

\renewcommand{\vec}[1]{\mbox{\boldmath $#1$}}

\begin{document}

\preprint{}

\title{Shape of $\Lambda$ hypernuclei in ($\beta,\gamma$) deformation plane}

\author{Myaing Thi Win}
\affiliation{
Department of Physics, Tohoku University,
Sendai 980-8578, Japan}

\author{K. Hagino}
\affiliation{
Department of Physics, Tohoku University,
Sendai 980-8578, Japan}

\author{T. Koike}
\affiliation{
Department of Physics, Tohoku University,
Sendai 980-8578, Japan}

\date{\today}

\begin{abstract}

We study the shape of $\Lambda$ hypernuclei in the full 
($\beta,\gamma$) deformation 
plane, including both  
axially symmetric and triaxial quadrupole 
deformations. 
To this end, 
we use 
the constrained Skyrme Hartree-Fock+BCS method 
on the three-dimensional 
Cartesian mesh. 
The potential energy surface is analyzed 
for carbon hypernuclei as well as for sd-shell hypernuclei 
such as $^{27,29}_{~~~~\Lambda}$Si and $^{25,27}_{~~~~\Lambda}$Mg. 
We show that the potential energy surface 
in the ($\beta,\gamma$) plane is similar to each other 
between the hypernuclei and the corresponding 
core nuclei, although 
the addition of $\Lambda$ hyperon makes 
the energy surface somewhat softer along the $\gamma$ direction. 
\end{abstract}


\maketitle

\section{Introduction}

One of the main interests in hypernuclear physics is to investigate 
how an addition of $\Lambda$ particle influences the properties 
of atomic nuclei. 
Due to the absence of Pauli's principle 
between nucleon and $\Lambda$ particle, 
it is believed that a $\Lambda$ hyperon can 
be treated as an {\it impurity} to probe deep interior of 
the nuclear medium. 
With the presence of hyperon as an impurity, 
some bulk properties of nuclei 
such as the shape and collective motions may be changed 
\cite{Motoba83,Hiyama99,Zofka80}. 
Indeed, 
the shrinkage of 
$^7_{\Lambda}$Li,  
with respect to 
$^6$Li,  has been observed 
experimentally by measuring 
the B(E2) value from the $5/2^+$ state to the $1/2^+$ state 
of $^7_{\Lambda}$Li 
\cite{Tanida01,Hashimoto06}. 

As the shape of nuclei plays a decisive role in determining their 
properties such as quadrupole moment and radius, 
mean-field model calculations have been performed in recent years  
to 
investigate the change of nuclear shape due to the addition of a $\Lambda$ 
hyperon. 
Deformed Skyrme-Hartree-Fock(SHF) studies in Ref. \cite {Zhou07} have shown 
that the deformation parameter of the hypernuclei which they 
studied is slightly smaller (within the same sign) than 
that of the corresponding 
core nuclei, and thus 
no significant effect of $\Lambda$ hyperon 
on nuclear deformation was found. 
On the other hand, we have 
performed 
a relativistic mean field (RMF) study 
and 
found that 
the nuclear deformation of 
the core nuclei completely 
disappears 
for $^{13}_{~\Lambda}$C and $^{29}_{~\Lambda}$Si hypernuclei 
due to the addition of $\Lambda$ particle \cite{Myaing08}, 
although we have obtained  
similar results to the SHF calculations in Ref. \cite{Zhou07} for 
many other hypernuclei. 
In Ref. \cite{Schulze10}, we have compared between the SHF and RMF 
approaches and have shown that the different results with 
respect to nuclear deformation between the two approaches is 
due to the fact that the RMF yields a somewhat stronger 
polarization effect of $\Lambda$ hyperon as compared to the SHF approach. 
We have also shown that the disappearance of the deformation 
realizes also in the SHF approach if the energy difference between 
the optimum deformation and the spherical configuration is less than 
about 1 MeV \cite{Schulze10}. 

All of these mean-field calculations 
have assumed axial symmetric deformation.  
Although many nuclei are considered to have axially symmetric shape, 
the triaxial degree of freedom plays an important role 
in transitional nuclei, nuclei 
with shape coexistence, 
and nuclei with gamma soft deformation 
\cite{Hayashi84,Redon86,Javid01,Bonche85,Girod09,ZPLi10,Kurath71}. 
In particular, we mention that 
recent studies with the constrained Hartree-Fock-Bogoliubov plus 
local quasi-particle random phase approximation (CHFB+LQRPA) 
method \cite{Hinohara10} 
as well as the RMF plus generator coordinate method (RMF+GCM) \cite{Yao} 
have revealed an important role of triaxiality in 
large amplitude collective motion in 
sd-shell nuclei. 

The aim of this paper is to extend the previous mean-field studies 
on deformation of hypernuclei 
by taking into
account the triaxial degree of freedom, that is, 
by including both the $\beta$ and
$\gamma$ deformations 
in order to investigate the effect of 
$\Lambda$ hyperon in the full ($\beta$,$\gamma$) deformation plane.  
We particularly study 
the potential energy surface (PES) in ($\beta$,$\gamma$) deformation plane 
with the SHF method 
for Carbon hypernuclei as well as some sd-shell hypernuclei.  
Notice that the shape evolution of the C isotopes 
in the full ($\beta,\gamma$) plane 
has been studied by Zhang {\it et al.} using the SHF method \cite{Ying08}. 
We extend the work of Zhang {\it et al.} by introducing a hyperon degree of 
freedom. 

The paper is organised as follows. 
In Sec. II, 
we briefly summarise the Skyrme-Hartree-Fock method 
for hypernuclei. 
In Sec. III, we present the results for the potential energy surface 
for the C isotopes as well as for 
sd-shell nuclei, 
$^{27,29}_{~~~~\Lambda}$Si and $^{25,27}_{~~~~\Lambda}$Mg. 
We also discuss the softness of the energy surface 
along the $\gamma$ deformation. 
In Sec. IV, we summarize the paper.

\section{Skyrme Hartree-Fock+BCS method for Hypernuclei}

The self-consistent mean field approach provides a useful 
means to study 
the ground state properties of hypernuclei. 
The core polarization effect, that is, 
the change of properties of a core nucleus due to an addition of 
$\Lambda$ particle such as 
a change of total energy and radius 
can be automatically taken into account with this method 
\cite{Lanskoy98,Harada05}. 
The self-consistent non-relativistic mean field calculations 
with a Skyrme-type $\Lambda$-nucleon($\Lambda$N) interaction 
have been performed by Rayet in Refs. \cite{Rayet76,Rayet81}. 
The relativistic mean field approach has also been applied to 
hypernuclei in {\it e.g.,} Refs. \cite{Rufa87,Mares89,Rufa90,Vretener98}. 
It has been pointed out that the neutron drip line 
is extended by the addition of $\Lambda$ hyperon \cite{Zhou08,Vretener98}.

In the present paper, we employ the Skyrme-type $\Lambda$N interaction 
and perform mean-field calculations 
by extending the computer code {\tt ev8}
\cite{Bonche05} to $\Lambda$ hypernuclei. 
The code solves the Hartree-Fock equations 
by discretizing individual single-particle 
wave functions on a three-dimensional Cartesian mesh. 
The pairing correlation is taken into account in the 
BCS approximation. 
With this method, 
both axial and triaxial quadrupole deformations can be 
automatically described. 
The code {\tt ev8} has been applied to the study 
of shape transition and deformation of 
several nuclei \cite{Sarriguren08,Tajima96,Bonche85,Ying08} in the  
($\beta,\gamma$) deformation plane. 

The Skyrme-type $\Lambda$N interaction 
is given in complete analogy with the nuclear Skyrme interaction 
\cite{Rayet81}. 
The Skyrme part of the total hypernuclear energy thus reads 
\begin{equation}
E= \int d^3r[H_N({\bf{r}})+H_\Lambda({\bf{r}})], 
\end{equation}
where $H_N$({\bf{r}}) is the standard nuclear Hamiltonian density 
based on the Skyrme interaction. See {\it e.g.,} Refs. \cite{Vautherin72,BHR03} 
for its explicit form. 
$H_\Lambda$({\bf{r}}) is the hyperon Hamiltonian density given in terms of 
the lambda and nucleon densities as (with a correction 
for the coefficient of the 
$\vec{\nabla}\rho_N\cdot\vec{\nabla}\rho_\Lambda$ term) \cite{Rayet81} 
\begin{eqnarray}
H_\Lambda &=& \frac{\hbar ^2}{2m_\Lambda}\tau _\Lambda
+t^\Lambda_0\left(1+\frac{1}{2}x^\Lambda_0\right)\rho _\Lambda\rho_N\nonumber\\
&&+\frac{1}{4}(t^\Lambda_1+t^\Lambda_2)(\tau_\Lambda\rho_N+\tau_N\rho_\Lambda)\nonumber\\
&&+\frac{1}{8}(3t^\Lambda_{1}-t^\Lambda_2)(\vec{\nabla}\rho_N\cdot\vec{\nabla}\rho_\Lambda)
+\frac{1}{4}t^\Lambda_3\rho_\Lambda(\rho^2_N+2\rho_n\rho_p)\nonumber\\
&&+\frac{1}{2}W_0^\Lambda(\vec{\nabla}\rho_N\cdot\vec{J}_\Lambda+\vec{\nabla}
\rho_\Lambda\cdot\vec{J}_N). 
\end{eqnarray}
Here, 
$\rho _\Lambda$, $\tau _\Lambda$, and $\vec{J}_\Lambda$ are 
the particle density, the kinetic energy density, and the spin density 
of the $\Lambda$ hyperon. These are expressed using 
the single-particle wave function $\phi _\Lambda$ for the $\Lambda$ particle. 
$\rho _N$, $\tau _N$, and $\vec{J}_N$ are the total densities for 
the nucleons. 
$t^\Lambda_0, t^\Lambda_1, t^\Lambda_2, t^\Lambda_3, x_0^\Lambda$, and 
$W_0^\Lambda$ are the Skyrme parameters for the $\Lambda$N interaction. 

The Hartree-Fock equations for the single-particle wave functions 
are obtained by taking variation of the energy $E$. 
The equation for the nucleons reads, 
\begin{equation}
\left[-\vec{\nabla}\cdot \frac {\hbar ^2}{2m^*_q(\vec{r})}\vec{\nabla} 
+V_q(\vec{r}) + 
U_{N}(\vec{r})\right]\phi _q = e_q\phi _q,  
\label{HF-nu} 
\end{equation}
where $q$ refers to protons or neutrons, while that for the $\Lambda$ particle 
reads 
 \begin{equation}
\left[-\vec{\nabla}\cdot\frac {\hbar ^2}{2m^*_{\Lambda}(\vec{r})}\vec{\nabla} + 
U_\Lambda(\vec{r})\right]\phi _{\Lambda} = e_{\Lambda}\phi _{\Lambda}.  
\end{equation}
Here, $e_q$ and $e_\Lambda$ are the single-particle energies and 
the 
effective mass 
for the hyperon is given by 
\begin{equation}
\frac{\hbar^2}{2m^*_{\Lambda}(\vec{r})} = \frac{\hbar ^2}{2m_{\Lambda}}
+\frac{1}{4}(t^{\Lambda}_1+t^{\Lambda}_2) \rho _N(\vec{r}).
\end{equation}
$V_q(\vec{r})$ is the single-particle potential originating from the 
nucleon-nucleon Skyrme interaction 
\cite{Vautherin72,BHR03}. 
$U_{\Lambda}(\vec{r})$  and $U_N(\vec{r})$ are the single-particle potentials 
originating from the $\Lambda$N interaction. 
These are expressed as \cite{Rayet81} 
\begin{eqnarray}
U_{\Lambda}(\vec{r}) &=& t^\Lambda_0\left(1+\frac{1}{2}x^\Lambda_0\right)\rho_{N}
+\frac{1}{4}\left(t^\Lambda_1+t^\Lambda_2\right)\tau_{N}\nonumber\\
&&-\frac{1}{8}\left(3t^\Lambda_1-t^\Lambda_2\right){\bf\nabla}^2\rho_{N}
+\frac{t^\Lambda_3}{4}\left(\rho^2_N+2\rho_n\rho_p\right)\nonumber\\
&&-\frac{1}{2}W^\Lambda_0\vec{\nabla}\cdot\vec{J}_{N}\nonumber\\
&&+\frac{1}{2}W^\Lambda_0\vec{\nabla}\rho_{N}\cdot
(-i\vec{\nabla}\times\vec{\sigma}),
\end{eqnarray}
and
\begin{eqnarray}
U_{N}(\vec{r}) &=& t^\Lambda_0\left(1+\frac{1}{2}x^\Lambda_0\right)\rho_{\Lambda}
+\frac{1}{4}\left(t^\Lambda_1+t^\Lambda_2\right)\tau_{\Lambda}\nonumber\\
&&-\frac{1}{8}\left(3t^\Lambda_1-t^\Lambda_2\right){\bf\nabla}^2\rho_{\Lambda}\nonumber\\
&&+\frac{t^\Lambda_3}{4}\rho_\Lambda(2\rho_N+2(\rho_N-\rho_q))
-\frac{1}{2}W^\Lambda_0\vec{\nabla}\cdot\vec{J}_{\Lambda}\nonumber\\
&&+\frac{1}{2}W^\Lambda_0\vec{\nabla}\rho_{\Lambda}\cdot
(-i\vec{\nabla}\times\vec{\sigma}).
\end{eqnarray}
The pairing correlations among the nucleons are treated in the 
BCS approximation. 
For the pairing interaction, we employ a zero-range 
density-dependent pairing force \cite{Terasaki96},
\begin{equation}
V({\bf r_1, r_2})=-g\,\frac{1- \hat{P}^\sigma}{2}\,\left(1-\frac{\rho(\bar{\vec{r}})}
{\rho _0}\right)\delta(\vec{r}_1-\vec{r}_2),
\end{equation}
where $\hat{P}^\sigma$ is the spin-exchange operator, $\rho _0=0.16$ fm$^{-3}$, 
and $\bar{\vec{r}}=(\vec{r}_1+\vec{r}_2)/2$. 

In this paper, we mainly use the SGII interaction \cite{Vangiai81} 
for the NN interaction, while 
the set No.$1$ in Ref. \cite{Yamamoto88} for the $\Lambda$N interaction. 
The latter interaction was constructed 
by fitting to the binding energy of 
$^{17}_{~\Lambda}$O, yielding 
the well depth for a $\Lambda$ particle, $D_\Lambda$= 29.38 MeV, in 
infinite nuclear matter. 
Notice that the spin-orbit strength $W_0^\Lambda$ in the $\Lambda N$  interaction 
is zero for the set No. 1 \cite{Yamamoto88}. 
For the pairing interaction, 
we follow Ref. \cite{Ying08} to use 
$g=410$ MeV$\cdot$fm$^3$ for both protons and neutrons for Carbon 
hypernuclei, while we follow Ref. 
\cite{Terasaki96} to use $g=1000$ MeV$\cdot$fm$^3$ for calculations of
 sd-shell hypernuclei. 
A smooth pairing energy cutoff of 5 MeV around the Fermi level 
is used \cite{Terasaki96}. 
We assume that the $\Lambda$ particle occupies the lowest single-particle 
state.

Since the primary purpose of this paper is to draw 
the potential energy surfaces (PES) 
of $\Lambda$ hypernuclei as a function of $\beta$ and $\gamma$ deformation 
parameters, the isoscalar quadrupole 
constraint is imposed on the total energy. 
We relate the deformation parameter $\beta$ for 
hypernuclei with the total quadrupole moment $Q$ using 
the equation 
\begin{equation}
\beta \approx \frac{\sqrt{5\pi }Q}{3(A_c+1)R^2_0},
\end{equation}
where $A_c=A-1$ is the mass number of the core nucleus 
and $R_0=1.2A_c^{1/3}$ fm is the radius of the hypernucleus. 

\section{Results}
\subsection {Carbon Hypernuclei}

\begin{figure}[htb]
\includegraphics[width=0.8\linewidth, clip]{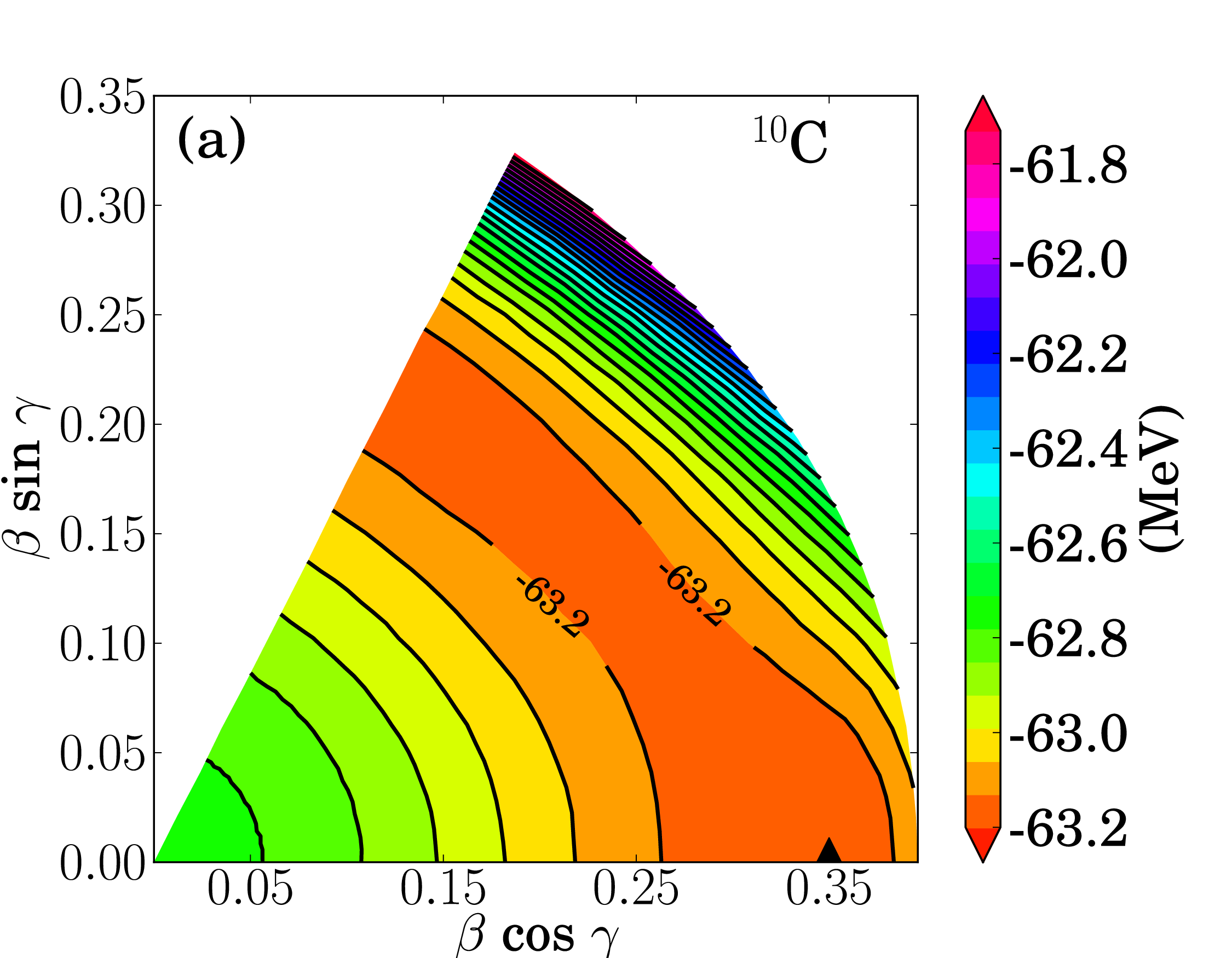}\\
\includegraphics[width=0.8\linewidth, clip]{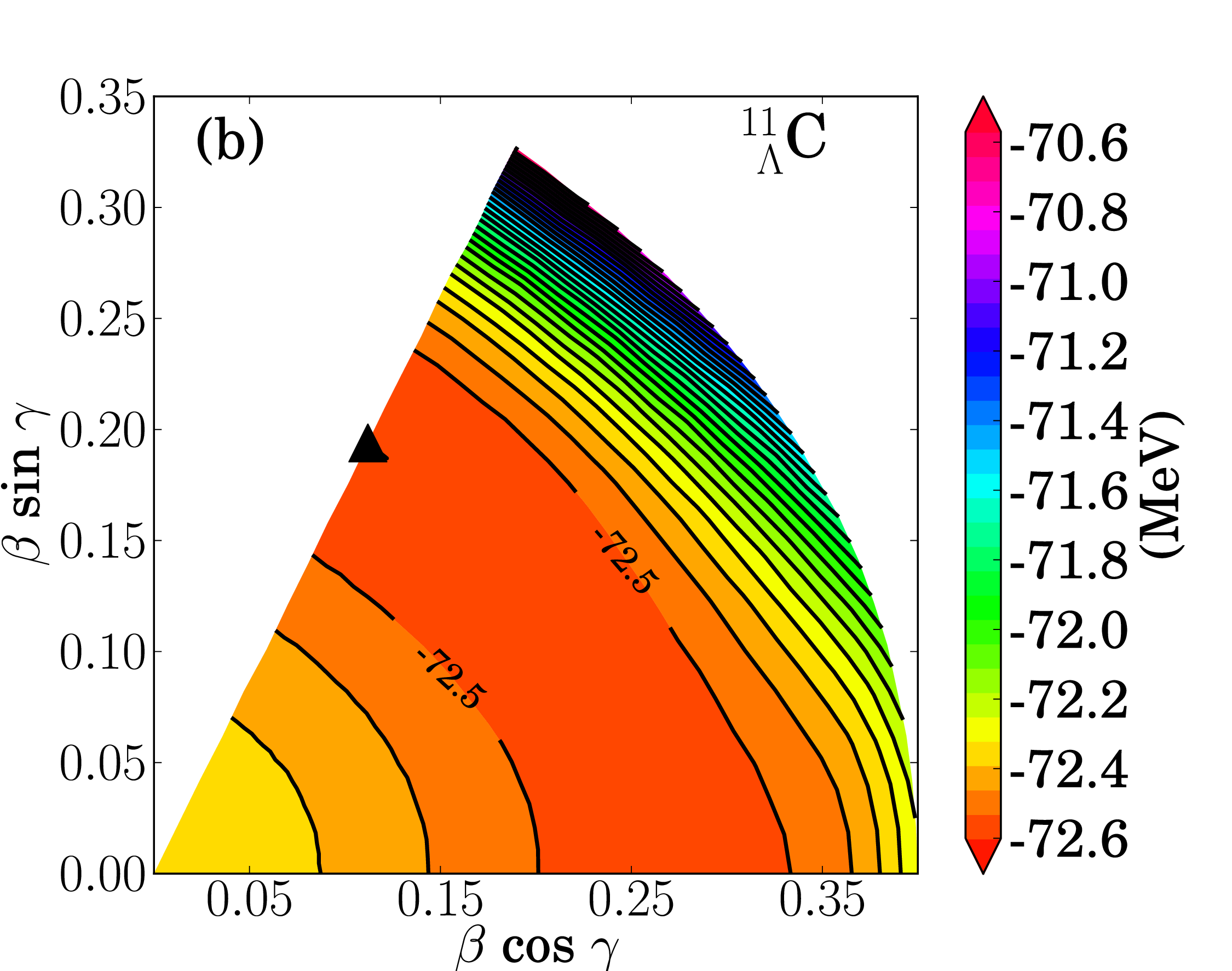} \\
\caption{(Color online) The potential energy surface (PES)  
of (a) $^{10}$C and (b) $^{11}_{~\Lambda}$C  
in the ($\beta,\gamma$) deformation plane 
obtained with the SGII parameter set. 
Each contour line is separated by 0.07MeV. 
The triangles indicate the absolute minima in the PES.  
}
\label{10C}
\end{figure}
\begin{figure}[htb]
\begin{center}\leavevmode
\includegraphics[width=0.7\linewidth, clip]{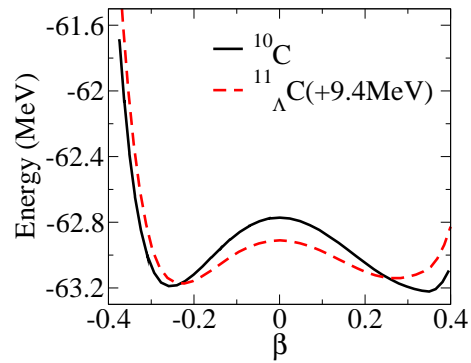}
\end{center}
\caption{The energy curve for $^{10}$C and $^{11}_{~\Lambda}$C
along the axially symmetric deformation corresponding to Fig. 1. 
The energy surface for $^{11}_{~\Lambda}$C
is shifted by a constant amount as indicated in the figure.}
\label{10Caxial}
\end{figure}

We now numerically solve the Hartree-Fock equations and discuss 
the deformation properties of hypernuclei
in the ($\beta,\gamma$) deformation plane. 
We first investigate 
the shape of carbon hypernuclei
from $^{11}_{~\Lambda}$C to $^{23}_{~\Lambda}$C. 
To this end, we follow Refs. \cite{Sagawa04,Ying08,Zhou07,Schulze10} 
and reduce the strength of the spin-orbit interaction 
by a factor of 0.6 
in the Skyrme functional. 
This prescription was introduced in order to reproduce 
an oblate ground state of $^{12}$C \cite{Sagawa04}. 

\begin{figure}[htb]
\begin{center}\leavevmode
\includegraphics[width=0.8\linewidth, clip]{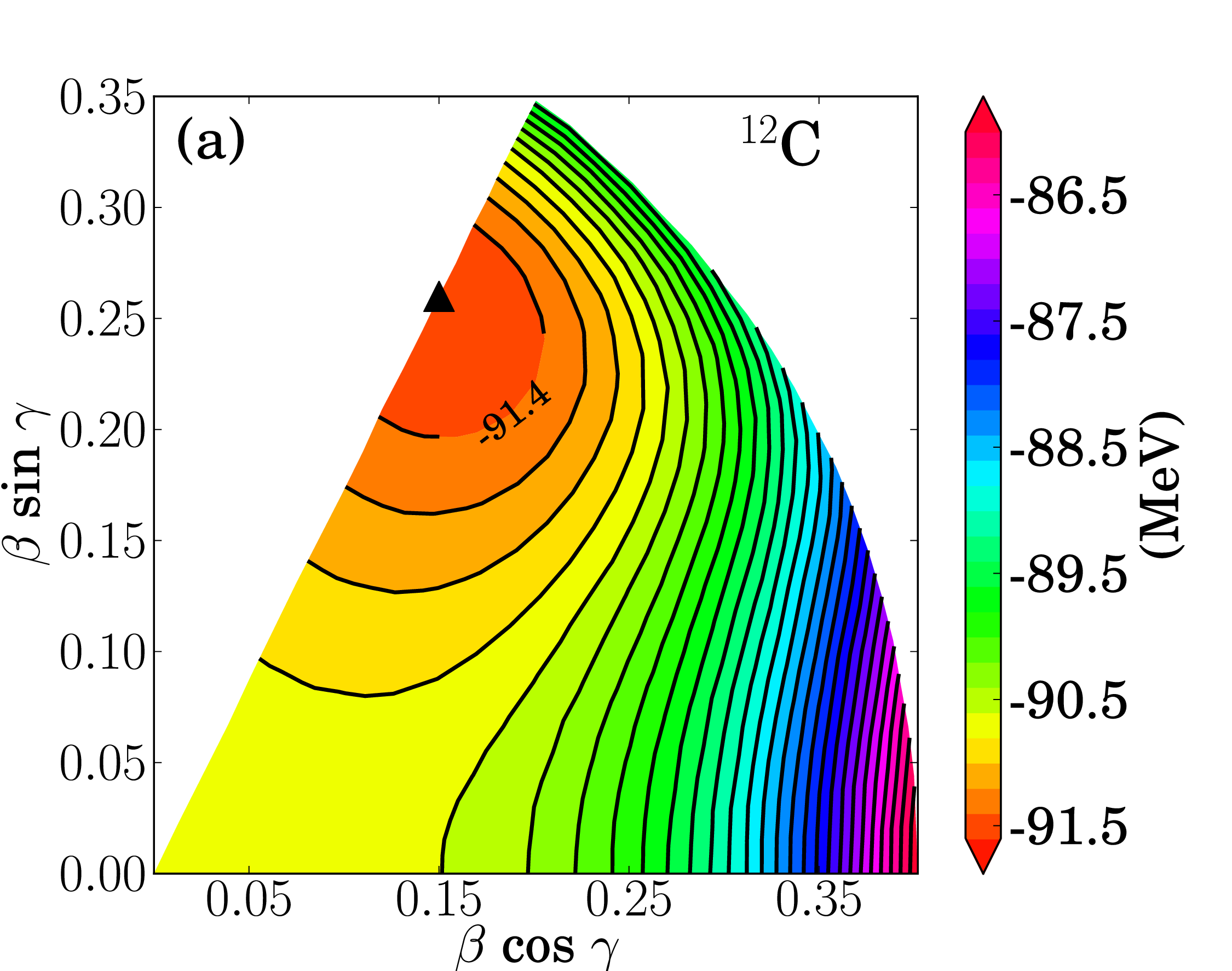}
\includegraphics[width=0.8\linewidth, clip]{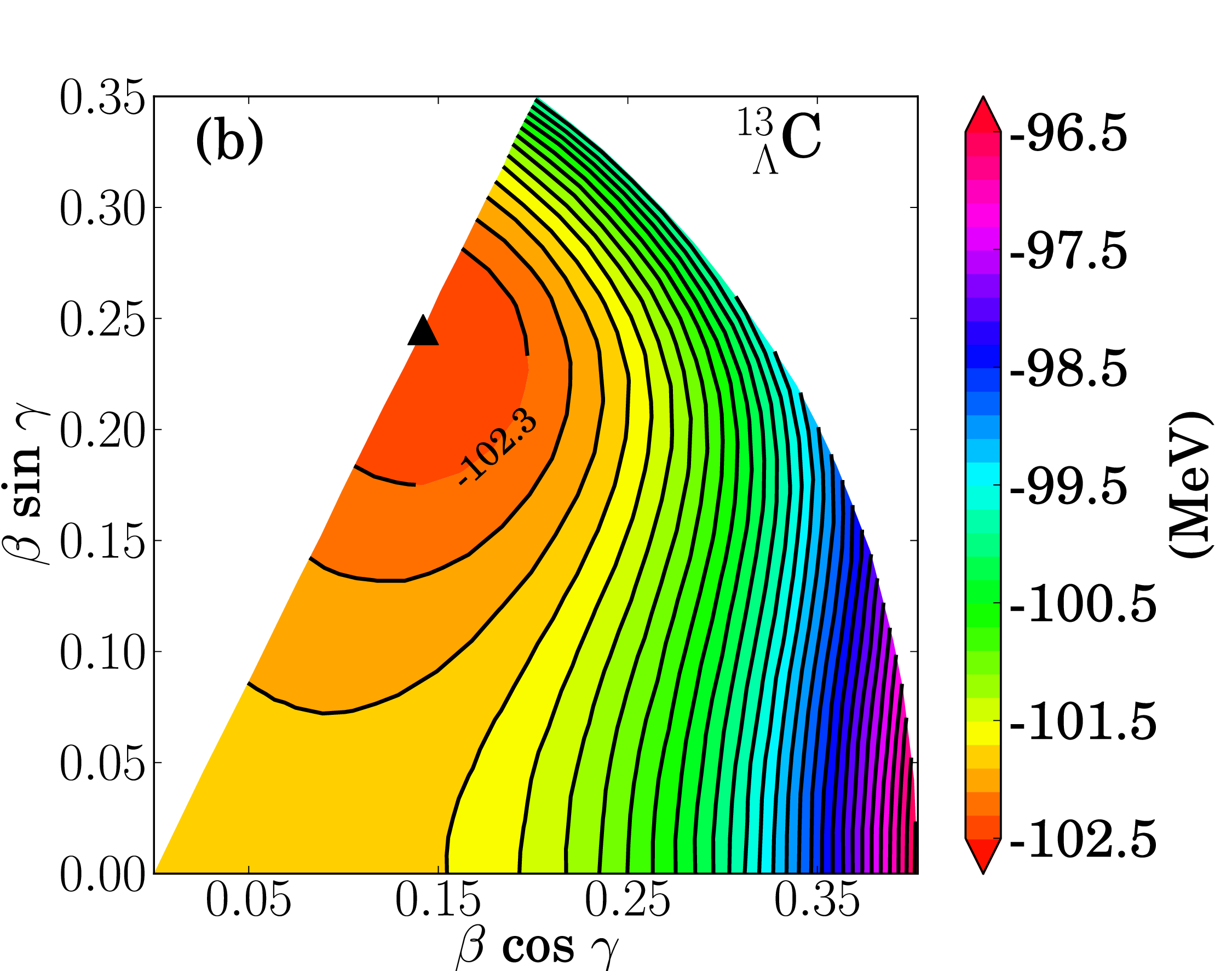}
\end{center}
\caption{(Color online) 
Same as Fig. 1, but for 
$^{12}$C and $^{13}_{~\Lambda}$C. 
Each contour line is separated by 0.2MeV.}
\label{12C}
\end{figure}
\begin{figure}[htb]
\begin{center}\leavevmode
\includegraphics[width=0.7\linewidth, clip]{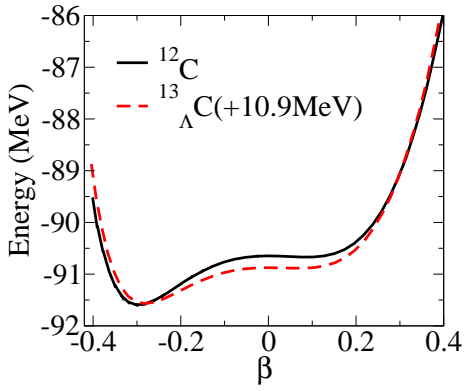}
\end{center}
\caption{(Color online) Same as Fig.2, but for 
$^{12}$C and $^{13}_{~\Lambda}$C. }
\label{12Caxial}
\end{figure}

Figure~\ref{10C} shows the 
potential energy surfaces for 
$^{10}$C and $^{11}_{~\Lambda}$C so obtained. 
The triangles in the figure indicate the 
ground state minimum in the energy surface. 
The energy curve along the 
axially symmetric deformation is 
also shown in Fig.~\ref{10Caxial} as a 
function of the quadrupole deformation parameter $\beta$. 
Along the axially symmetric configuration, 
there are deep energy minima in both sides of prolate and 
oblate configurations of $^{10}$C (that is, the shape coexistence), 
having a very small energy difference 
of less than 40 keV between them. 
The ground state corresponds to the prolate configuration 
with $\beta$=0.35. 
However, 
the energy surface is almost flat along the triaxial deformation $\gamma$ 
as one can see in Fig.~\ref{10C} (a) in the 
($\beta,\gamma$) deformation plane. 
With the addition of a $\Lambda$ hyperon, 
the ground state configuration moves from prolate to oblate, 
although the energy surface is so flat along the $\gamma$ degree of freedom 
(see Fig.~\ref{10C}(b)) that the ground state configuration 
may not be well defined in the mean field approximation. 
We have confirmed that this feature remains the same even if we use 
the Skyrme $\Lambda$N 
interaction set No.2 and No. 5 \cite{Yamamoto88}, instead of No. 1. 

\begin{figure}[htb]
\includegraphics[width=0.8\linewidth, clip]{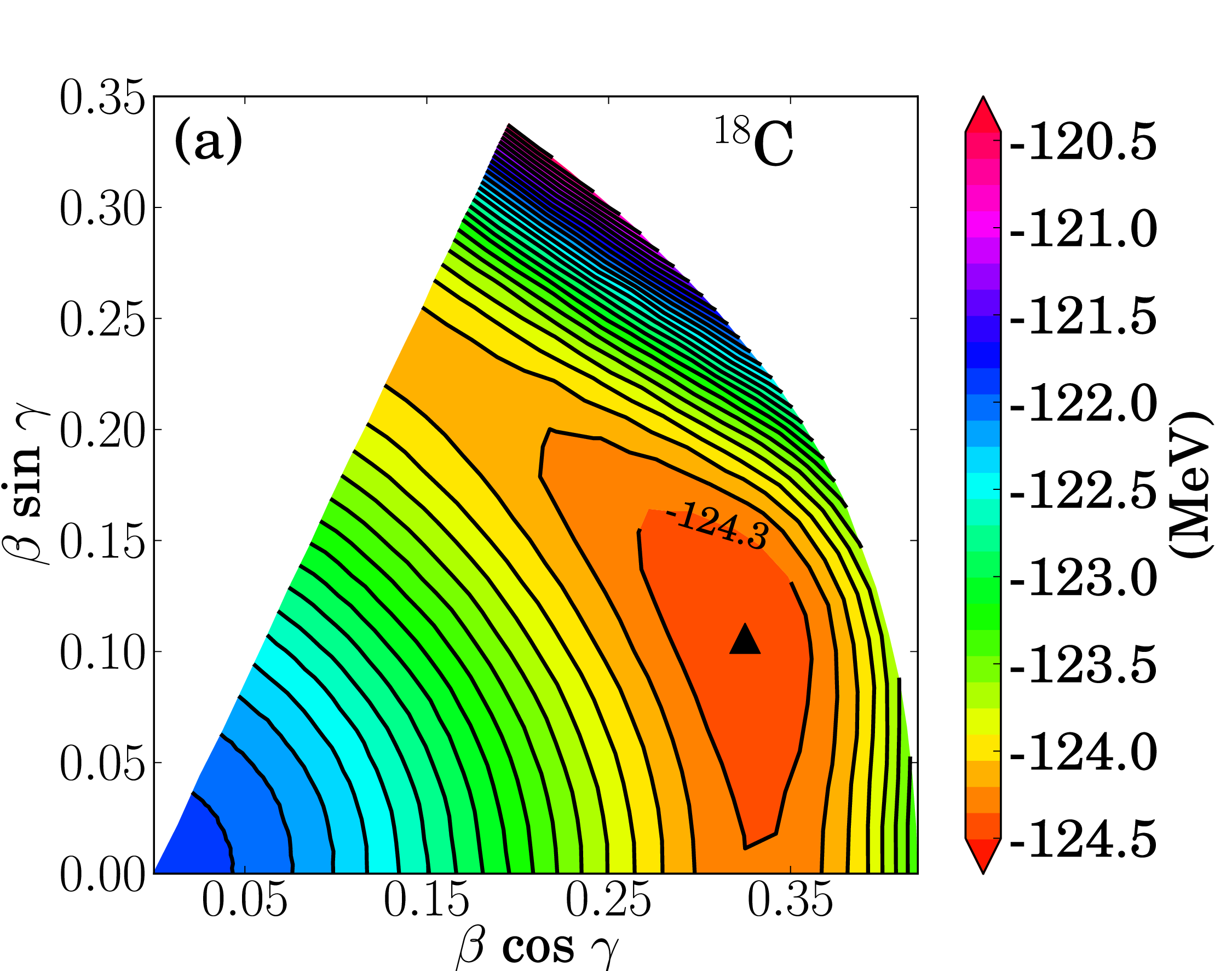}
\includegraphics[width=0.8\linewidth, clip]{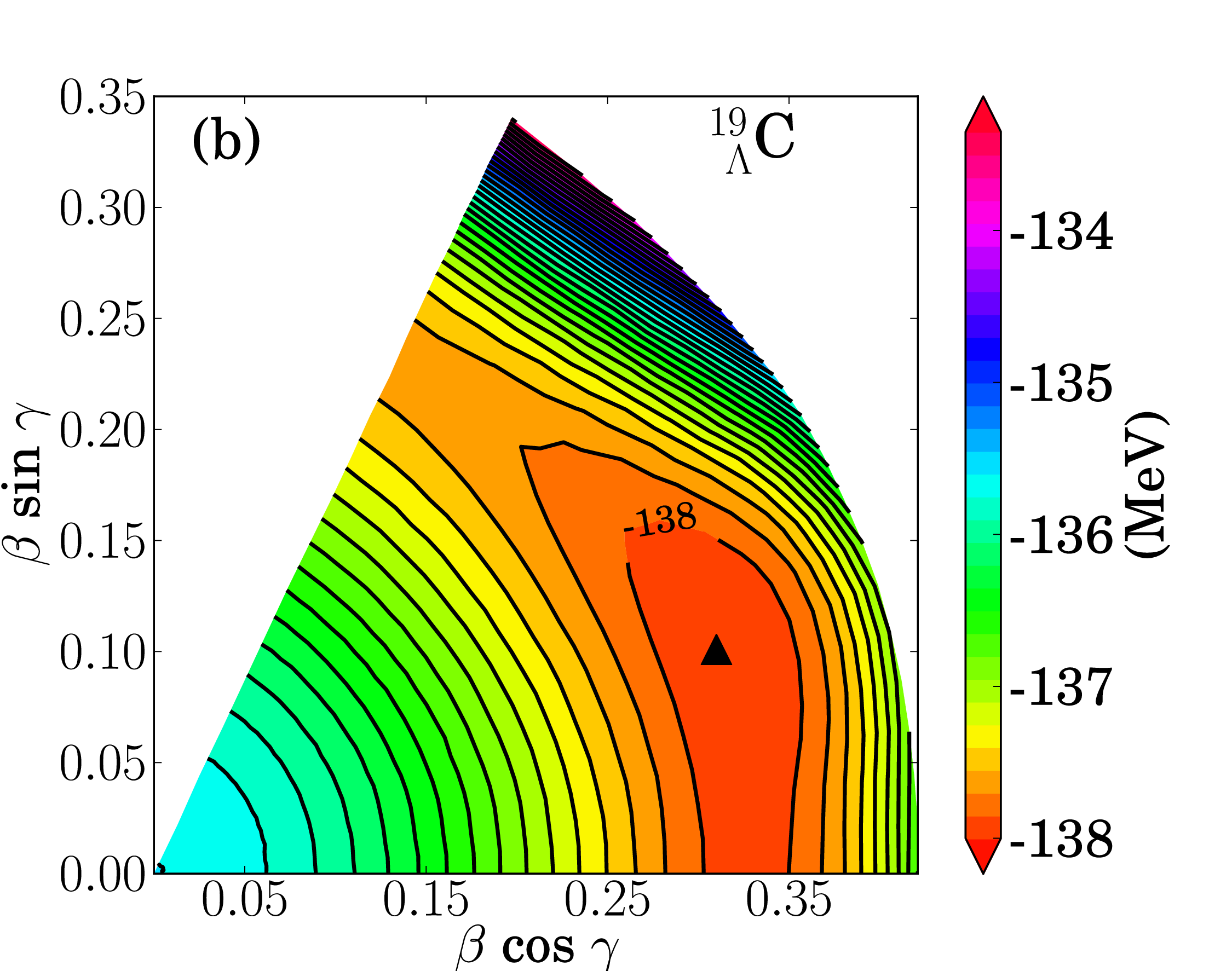}
\caption{(Color online) 
Same as Fig. 1, but for $^{18}$C and $^{19}_{~\Lambda}$C. 
Each contour line is separated by 0.15MeV} 
\label{18C}
\end{figure}
\begin{figure}[htb]
\begin{center}\leavevmode
\includegraphics[width=0.7\linewidth, clip]{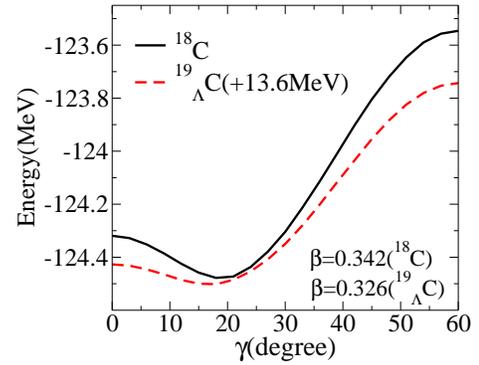}
\end{center}
\caption{(Color online) The energy curve as a function of 
the triaxial deformation parameter $\gamma$ for the optimum 
values of the $\beta$ deformation parameters shown in Fig. 
\ref{18C}. 
The energy surface for $^{19}_{~\Lambda}$C is shifted by 
a constant amount as indicated in the figure.}
\label{18C-g}
\end{figure}

Figs. \ref{12C} (a) and 
\ref{12Caxial} show the energy surface 
for $^{12}$C in the $(\beta,\gamma)$ deformation plane and that along the 
$(\beta,\gamma=0)$ line, respectively. 
For this nucleus, 
the Skyrme-Hartree-Fock method with the reduced spin-orbit interaction 
yields a deep oblate minimum. 
In this case, the addition of a $\Lambda$ particle shows little effect on 
the energy surface, although the deformation is slightly smaller than 
its core nucleus (see Figs.~\ref{12C}(b) and \ref{12Caxial}).
This is similar to the result for 
$^{20}$Ne discussed in Ref.\cite{Myaing08} with the RMF method.

For the case of $^{18}$C, it has a triaxial minimum at 
$\beta=0.34$ and $\gamma=18.0 ^{\circ}$ 
at an energy of $-$0.154MeV with respect to 
the prolate configuration (see Fig. \ref{18C} (a)). 
We find that 
$^{19}_{~\Lambda}$C has a ground state configuration with similar values 
of $\beta$ and $\gamma$ deformation parameters to those of the core nucleus,
$^{18}$C, as indicated in Fig. \ref{18C} (b). 
The triaxial minimum is again shallow 
with an energy difference of 0.077 MeV between the optimum deformation and the 
prolate configuration. 
In Fig.~\ref{18C-g}, we plot the energy curve 
as a function of the triaxial deformation parameter $\gamma$ 
for the optimum values of the $\beta$ deformation parameter. 
With the addition of a $\Lambda$ particle to $^{18}$C,
one sees that the energy curve becomes slightly softer 
(compare also Figs. \ref{18C}(a) and 
\ref{18C}(b)). 

\begin{figure}[htb]
\includegraphics[width=0.8\linewidth, clip]{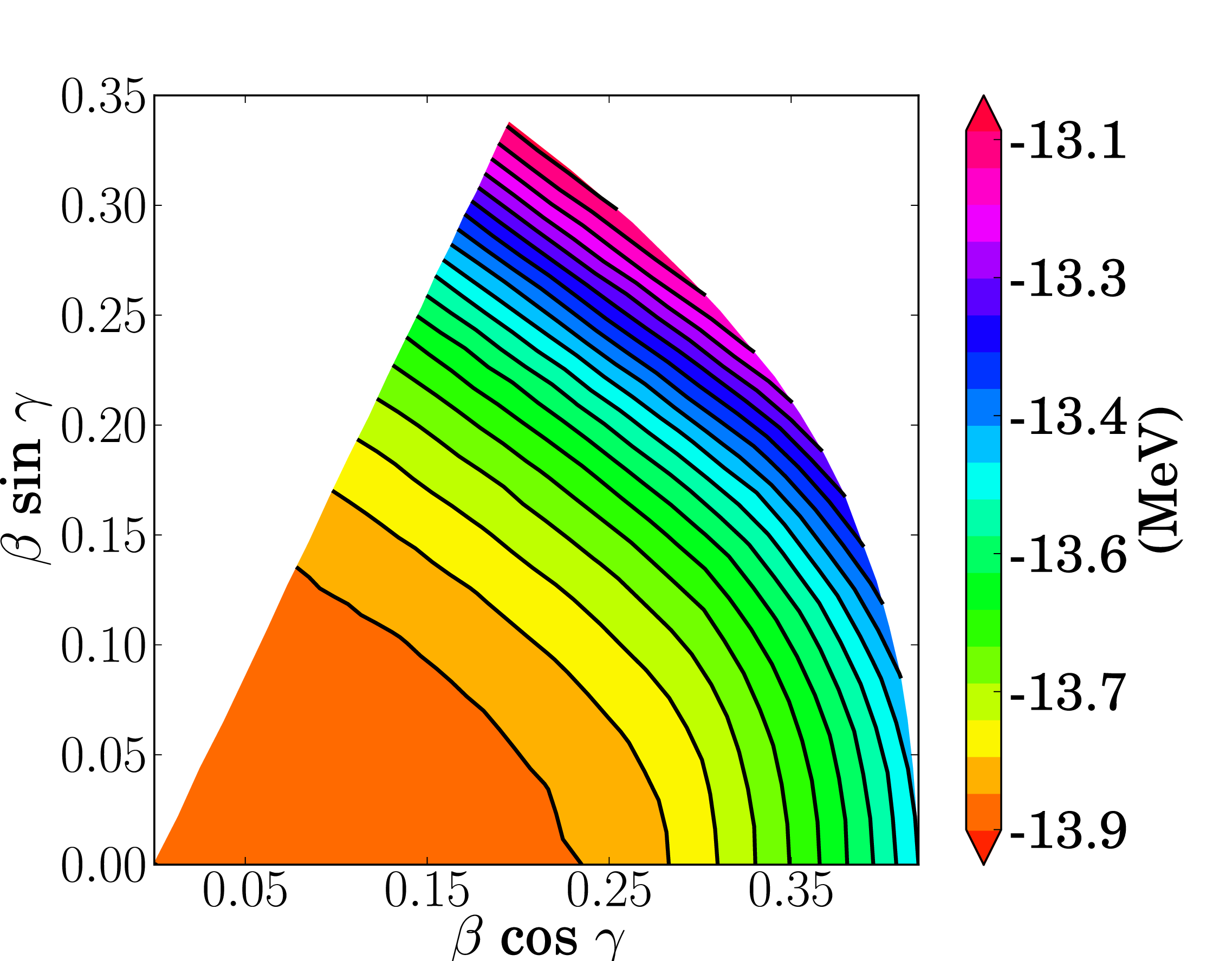}
\caption{(Color online) 
The energy difference between 
$^{19}_{~\Lambda}$C and $^{18}$C nuclei in the $(\beta,\gamma)$ deformation 
plane.}
\label{18CE}
\end{figure}

The energy difference between 
$^{19}_{~\Lambda}$C and $^{18}$C at each deformation, 
that is, 
$E_{^{19}_{~\Lambda}{\rm C}}(\beta,\gamma)-E_{^{18}{\rm C}}(\beta,\gamma)$, 
in the $(\beta,\gamma)$ deformation plane is plotted in Fig. \ref{18CE}. 
It clearly shows that 
the addition of a $\Lambda$ particle prefers the spherical configuration 
even if the core nucleus has a deformed minimum. 
We have confirmed that this is the case for the other carbon isotopes as well. 
Notice that the $\Lambda$ particle slightly prefers the prolate configuration 
for a fixed value of $\beta$, causing the slightly 
softer energy curve towards the prolate configuration shown 
in Fig. 5. This originates from the fact that the overlap between 
the deformed nuclear density and a spherical $\Lambda$ density 
is maximum at the prolate configuration, as 
we discuss in the Appendix.

The results of our calculations are summarised in Table I, together with 
the results for the other carbon isotopes. 

%
%
%

\subsection {sd shell Hypernuclei}

\subsubsection{Potential Energy Surface}

\begin{figure}[htb]
\begin{center}\leavevmode
\includegraphics[width=0.8\linewidth, clip]{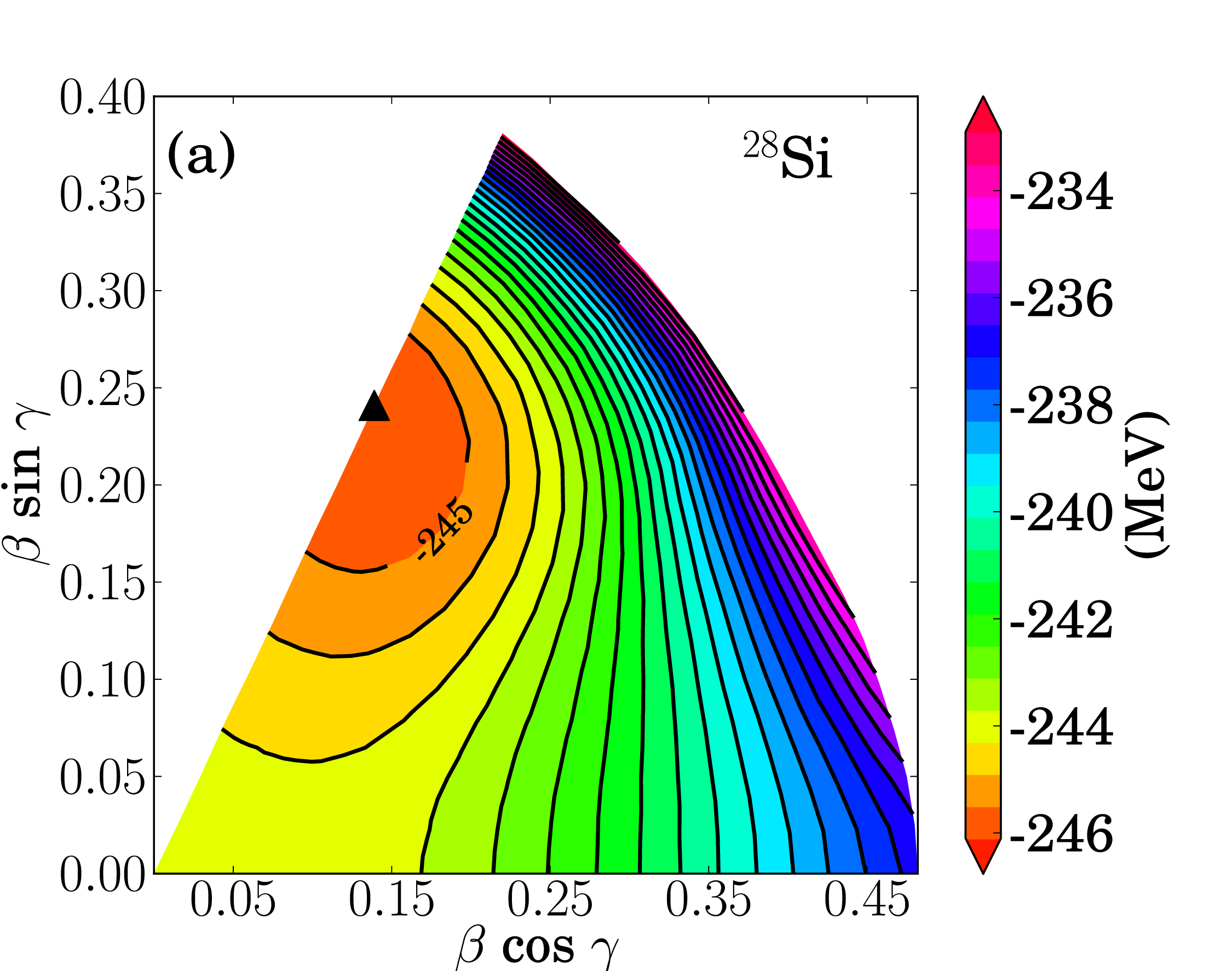}
\includegraphics[width=0.8\linewidth, clip]{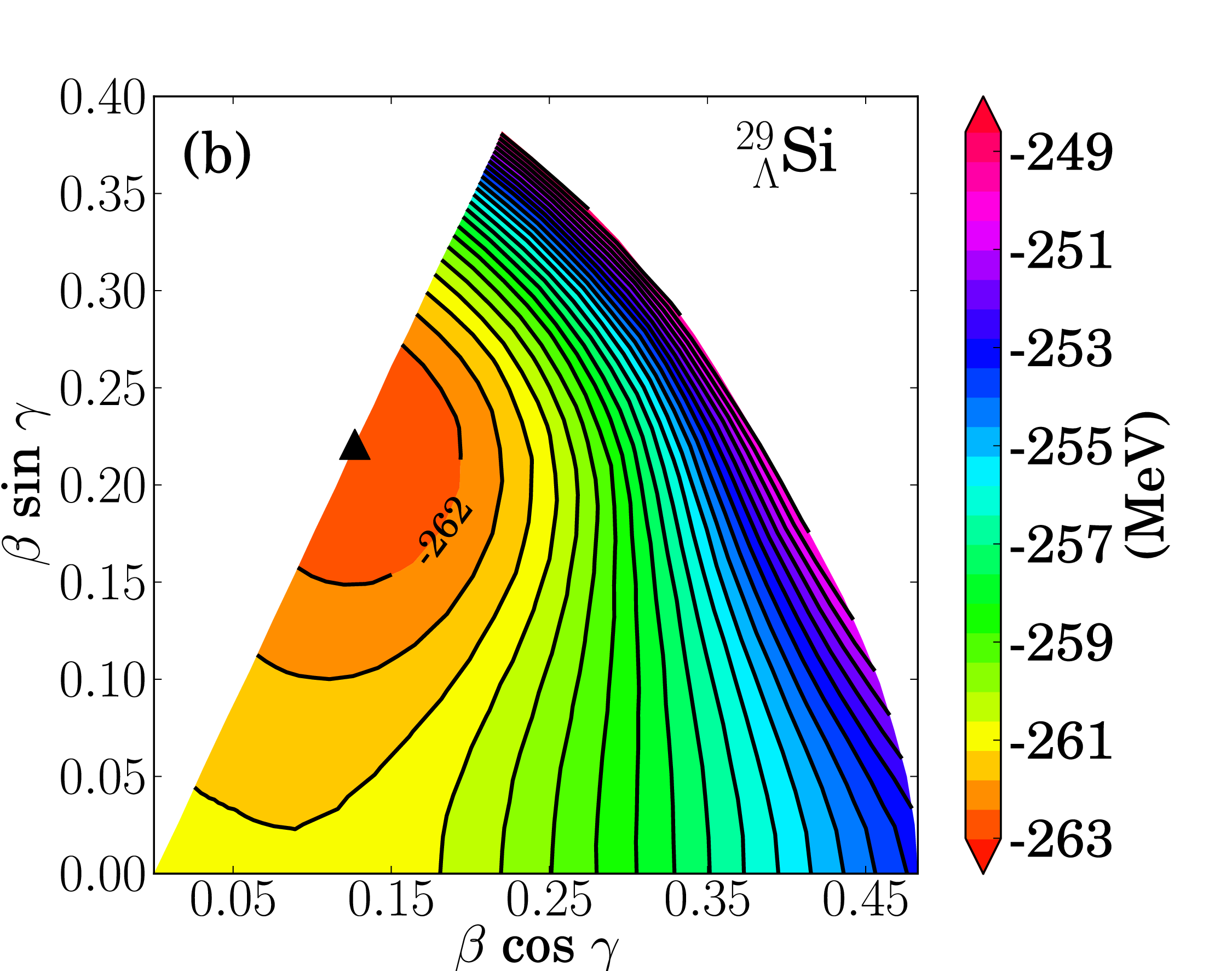}
\end{center}
\caption{(Color online) Same as Fig. 1, but for $^{28}$Si 
and $^{29}_{~\Lambda}$Si. Each contour line is separated by 0.6 MeV.}
\label{28Si}
\end{figure}
\begin{figure}[htb]
\begin{center}\leavevmode
\includegraphics[width=0.8\linewidth, clip]{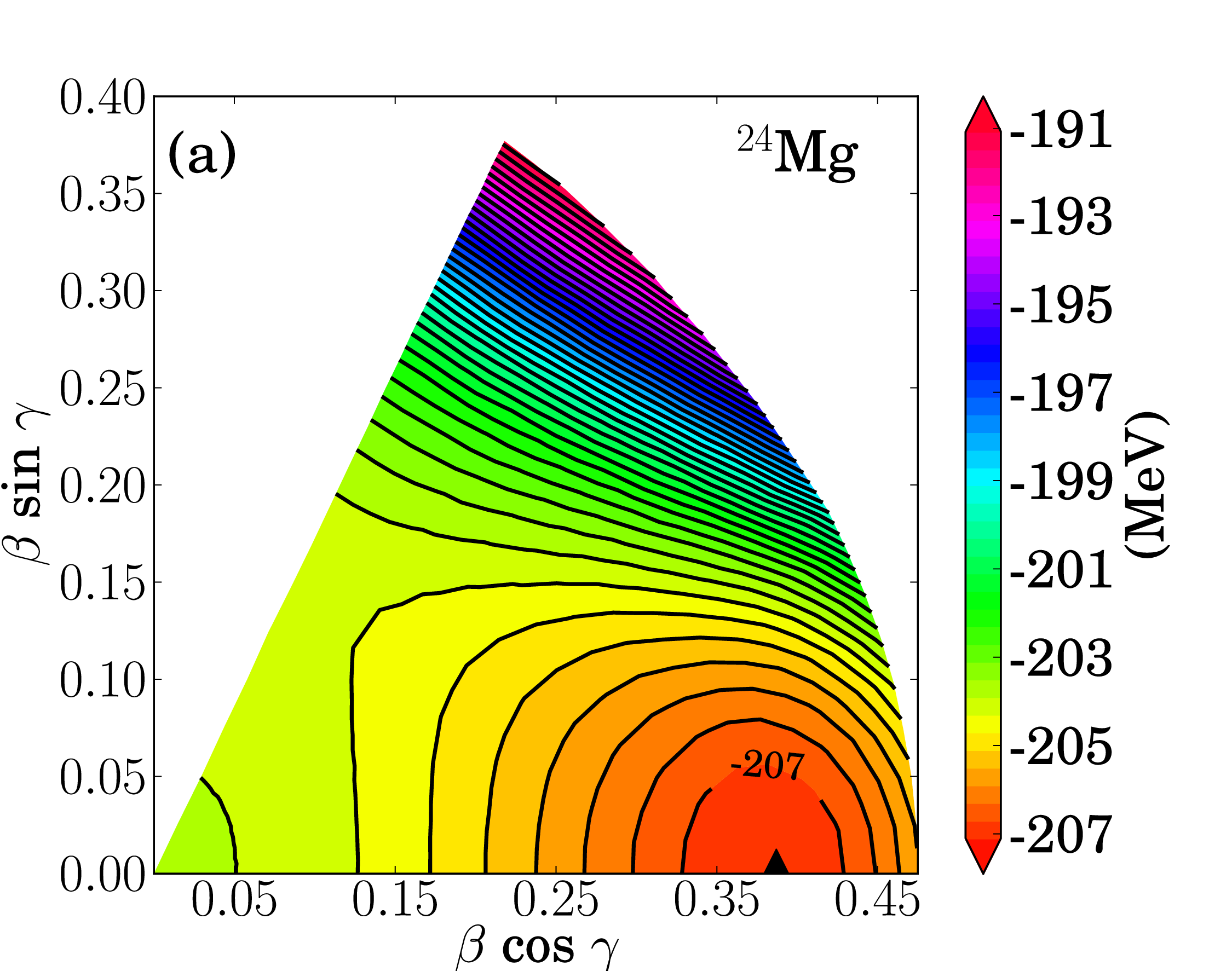}
\includegraphics[width=0.8\linewidth, clip]{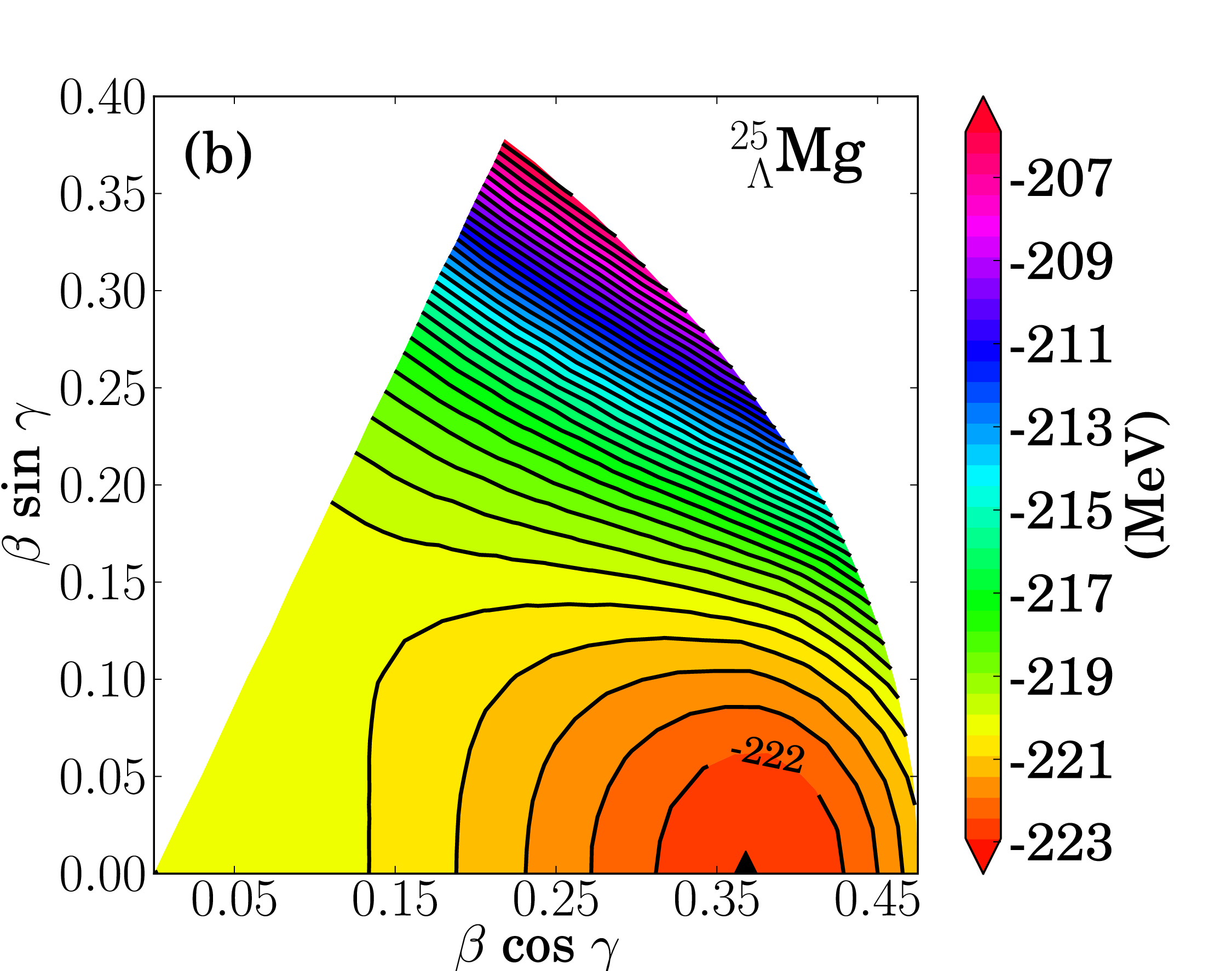}
\end{center}
\caption{(Color online) Same as Fig. 1, but for $^{24}$Mg
and $^{25}_{~\Lambda}$Mg. Each contour line is separated by 0.5 MeV.}
\label{24Mg}
\end{figure}

\begin{figure}[htb]
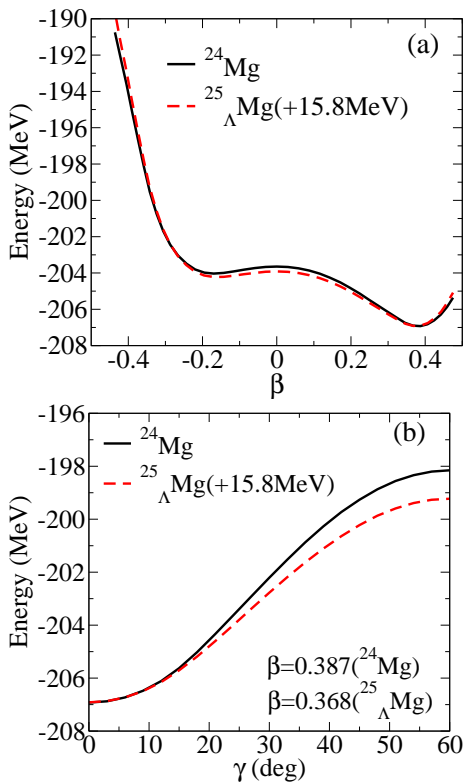

\begin{center}\leavevmode
\includegraphics[width=0.7\linewidth, clip]{fig10a.eps}
\includegraphics[width=0.7\linewidth, clip]{fig10b.eps}
\end{center}
\caption{(Color online) (a) The energy curve for $^{24}$Mg and $^{25}_{~\Lambda}$Mg 
along the axially symmetric deformation corresponding to Fig. 9. 
(b) The energy curve as a function of the 
triaxial deformation parameter $\gamma$ for the optimum values of 
the $\beta$ deformation parameters shown in Fig. 9. The energy 
surfaces for  $^{25}_{~\Lambda}$Mg are shifted by a constant amount as
indicated in the figures.}
\label{24axial}
\end{figure}

Let us next discuss the deformation energy 
surfaces of $^{27,29}_{~~~~\Lambda}$Si,and $^{25,27}_{~~~~\Lambda}$Mg 
nuclei in the ($\beta,\gamma$) deformation plane. 
For these nuclei, we use the original strength for the spin-orbit 
interaction. 
We first show the potential energy surface of 
$^{28}$Si and $^{24}$Mg in Figs. 
~\ref{28Si}(a) and Fig.~\ref{24Mg}(a), respectively. 
The energy surface for $^{28}$Si shows a deep oblate minimum, while 
that for $^{24}$Mg shows a deep prolate minimum. 
Notice that N=Z=14 is an oblate magic number. 
The energy curves for $^{24}$Mg 
as a function of $\beta$ with $\gamma=0$, and 
of $\gamma$ with $\beta=\beta_{\rm min}=0.387$ 
are also plotted in 
in Figs.~\ref{24axial}(a) and ~\ref{24axial}(b), respectively. 

For these nuclei, as the potential minimum is deep, 
the addition of a $\Lambda$ particle does not change 
significantly the shape of the potential energy surface, as shown 
in Figs. ~\ref{28Si}(b),~\ref{24Mg}(b),and ~\ref{24axial}. 
As shown in the previous subsection, the energy gain due to 
the additional $\Lambda$ particle appears 
mainly around the spherical configuration.  
Notice, however, that the $\Lambda$ particle makes 
the energy curve slightly softer along the triaxial degree of freedom, 
$\gamma$. 

\begin{figure}[htb]
\begin{center}\leavevmode
\includegraphics[width=0.8\linewidth, clip]{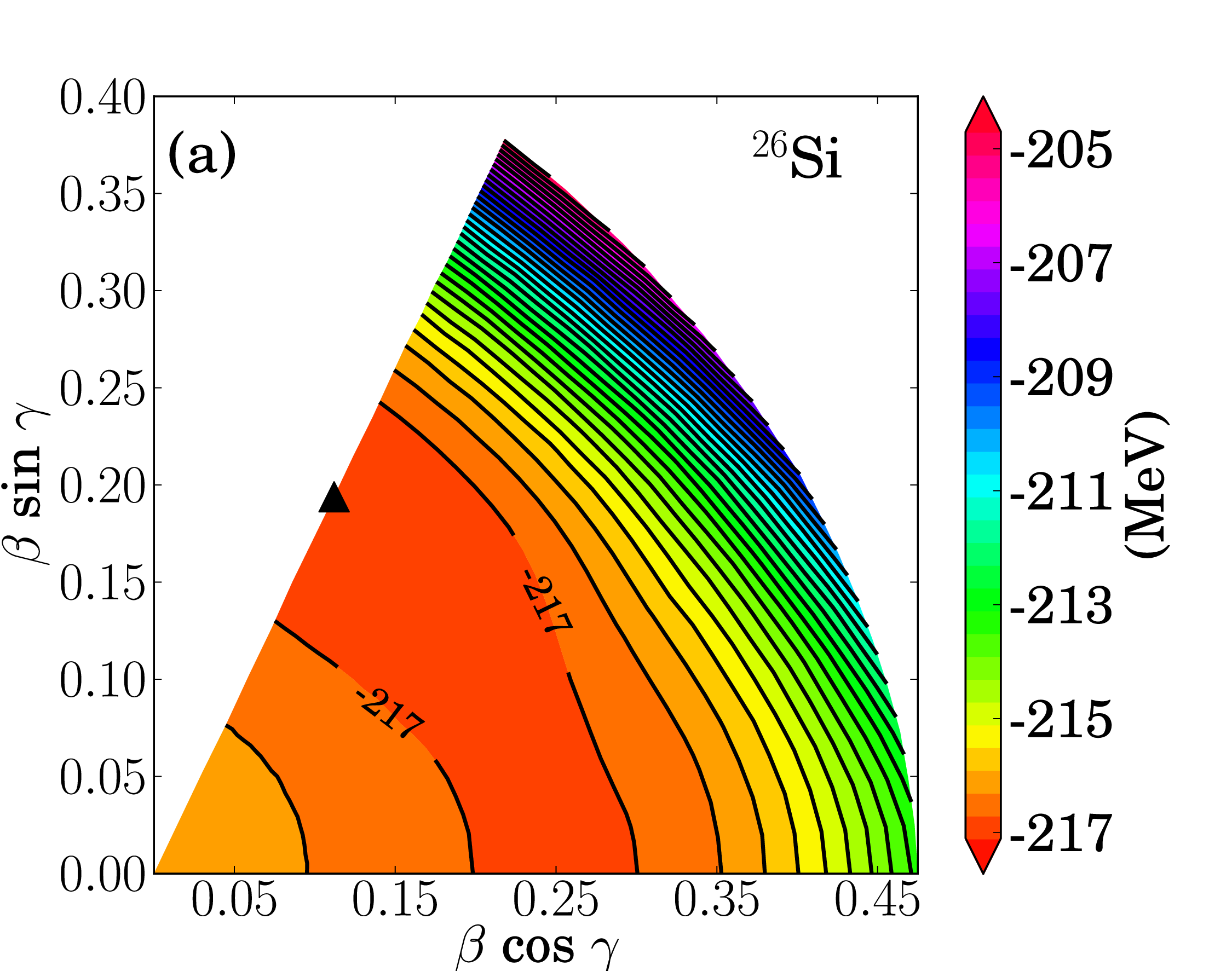}
\includegraphics[width=0.8\linewidth, clip]{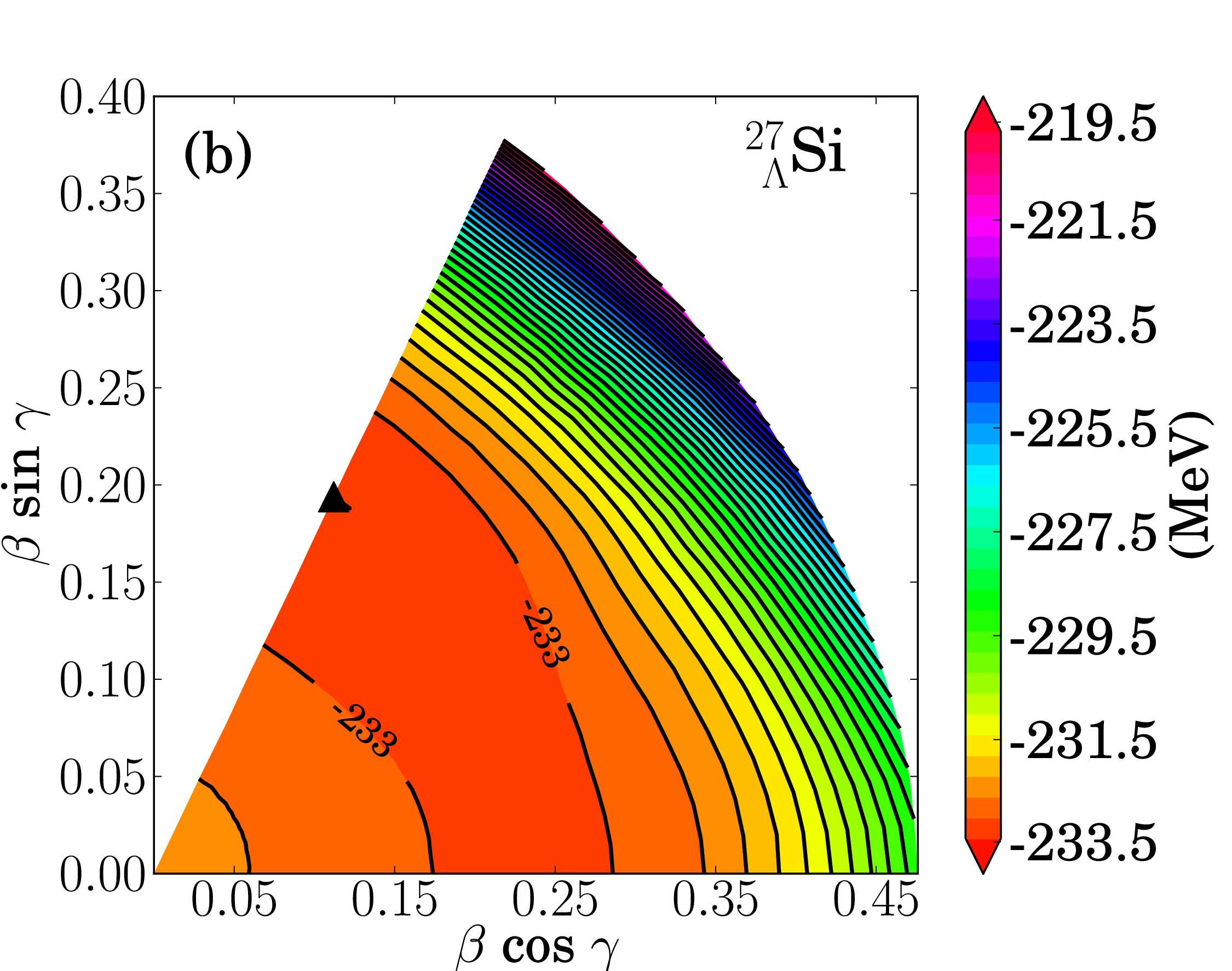}
\end{center}
\caption{(Color online) Same as Fig. 1, but for 
 $^{26}$Si and $^{27}_{~\Lambda}$Si.  
 Each contour line is separated by 0.4 MeV.}
\label{26Si}
\end{figure}

\begin{figure}[htb]
\begin{center}\leavevmode
\includegraphics[width=0.8\linewidth, clip]{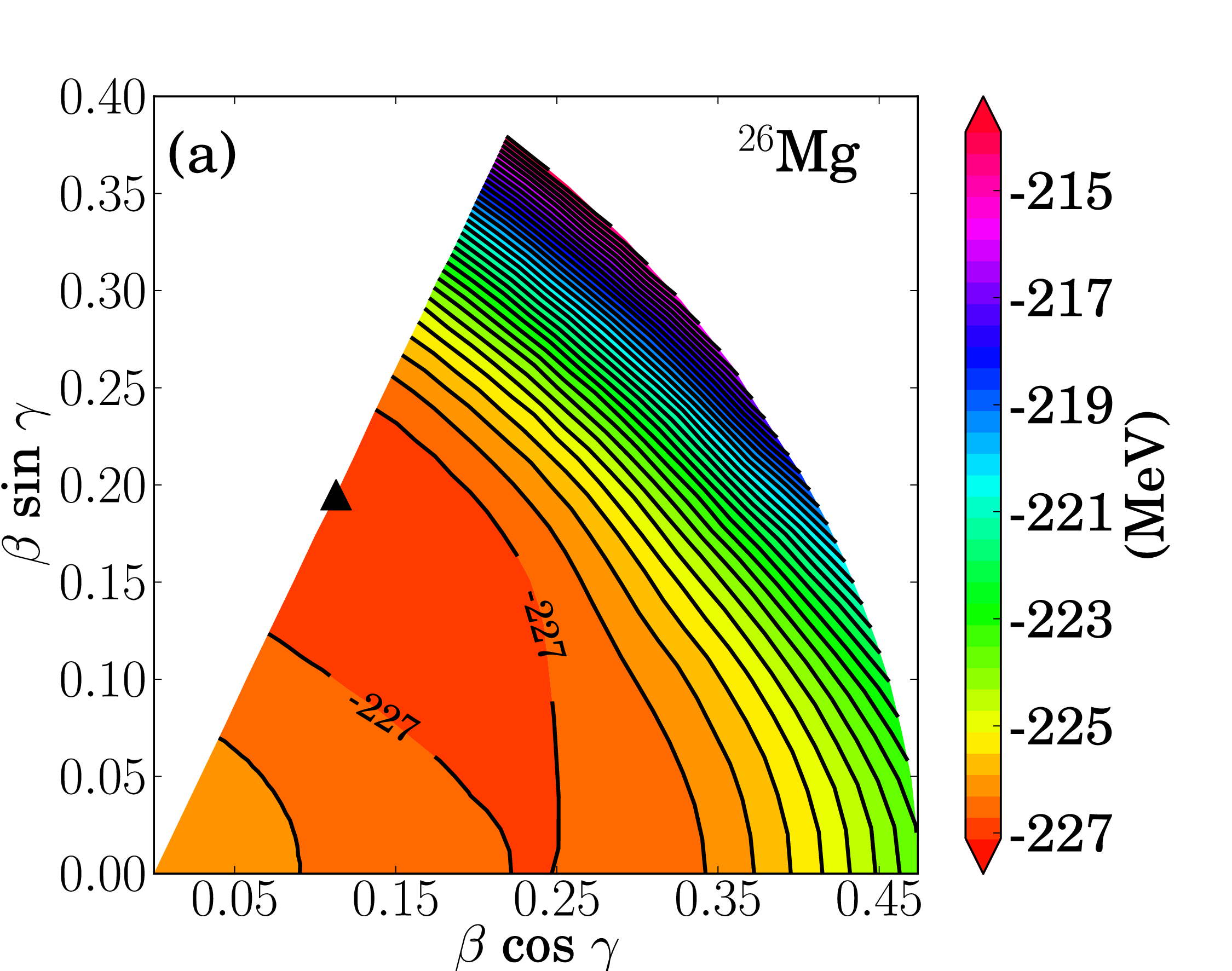}
\includegraphics[width=0.8\linewidth, clip]{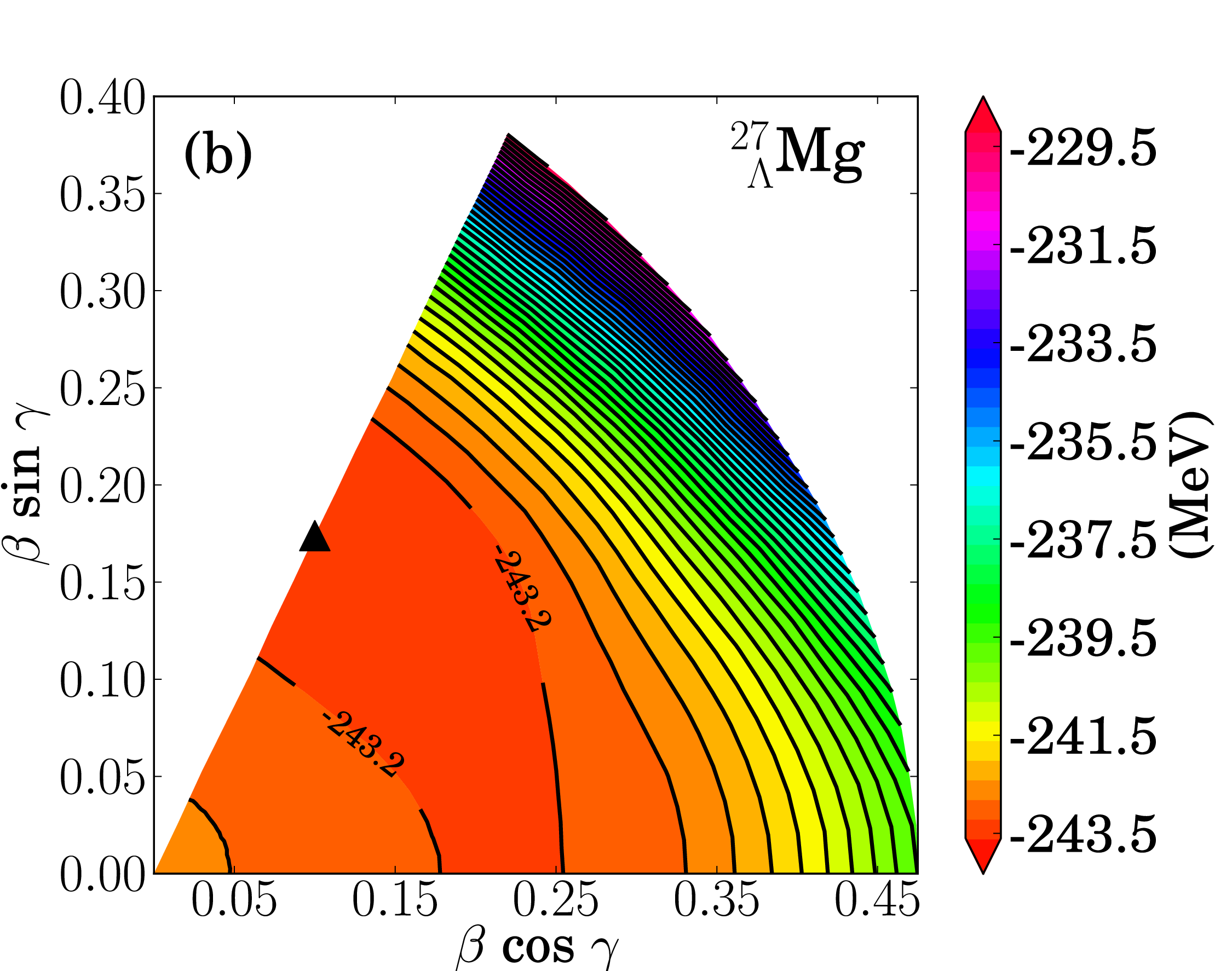}
\end{center}
\caption{(Color online)  Same as Fig. 1, but for $^{26}$Mg and
 $^{27}_{~\Lambda}$Mg. Each contour line is separated by 0.4 MeV.}
\label{26Mg}
\end{figure}

\begin{figure}[htb]
\begin{center}\leavevmode
\includegraphics[width=0.7\linewidth, clip]{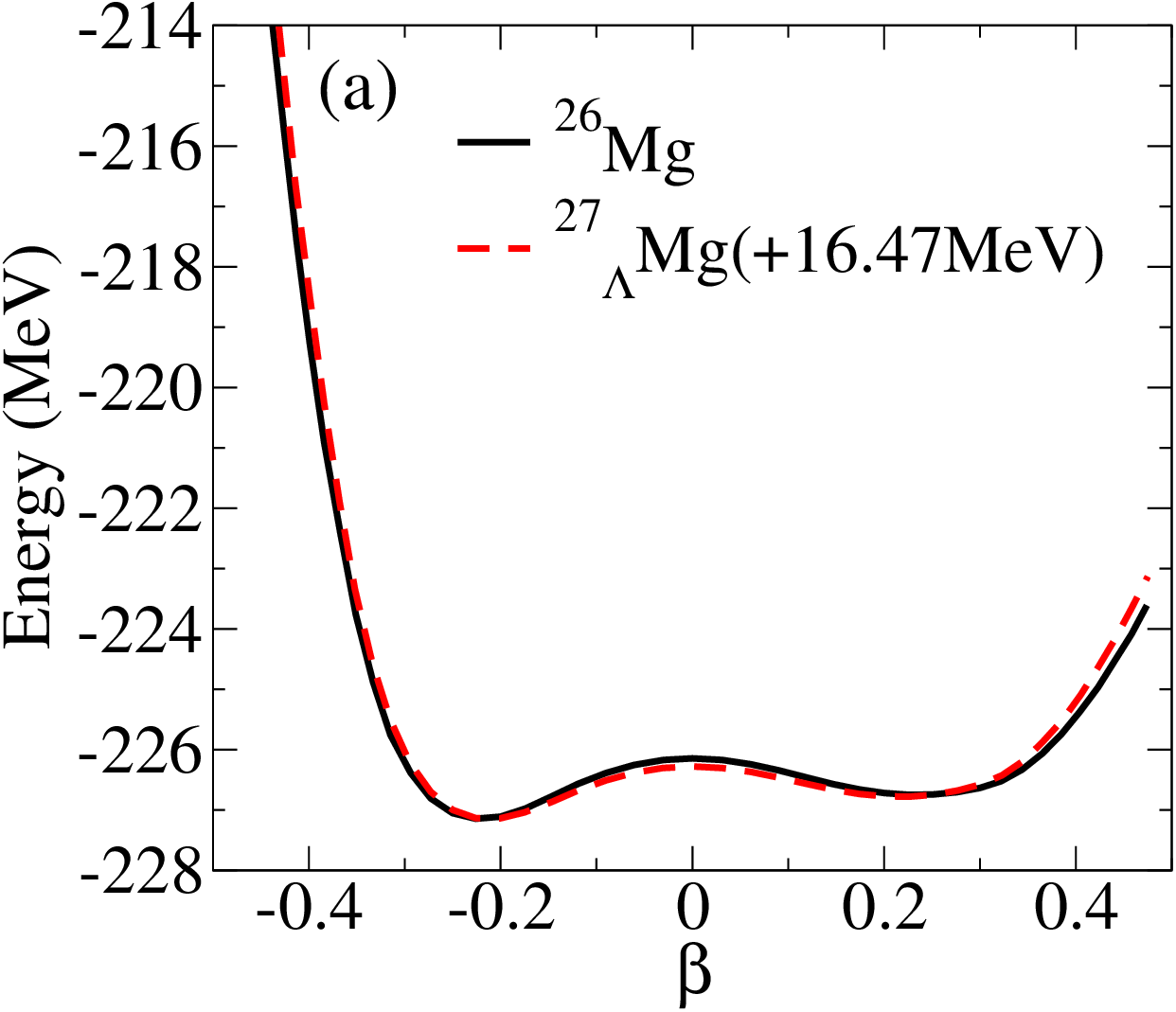}
\includegraphics[width=0.7\linewidth, clip]{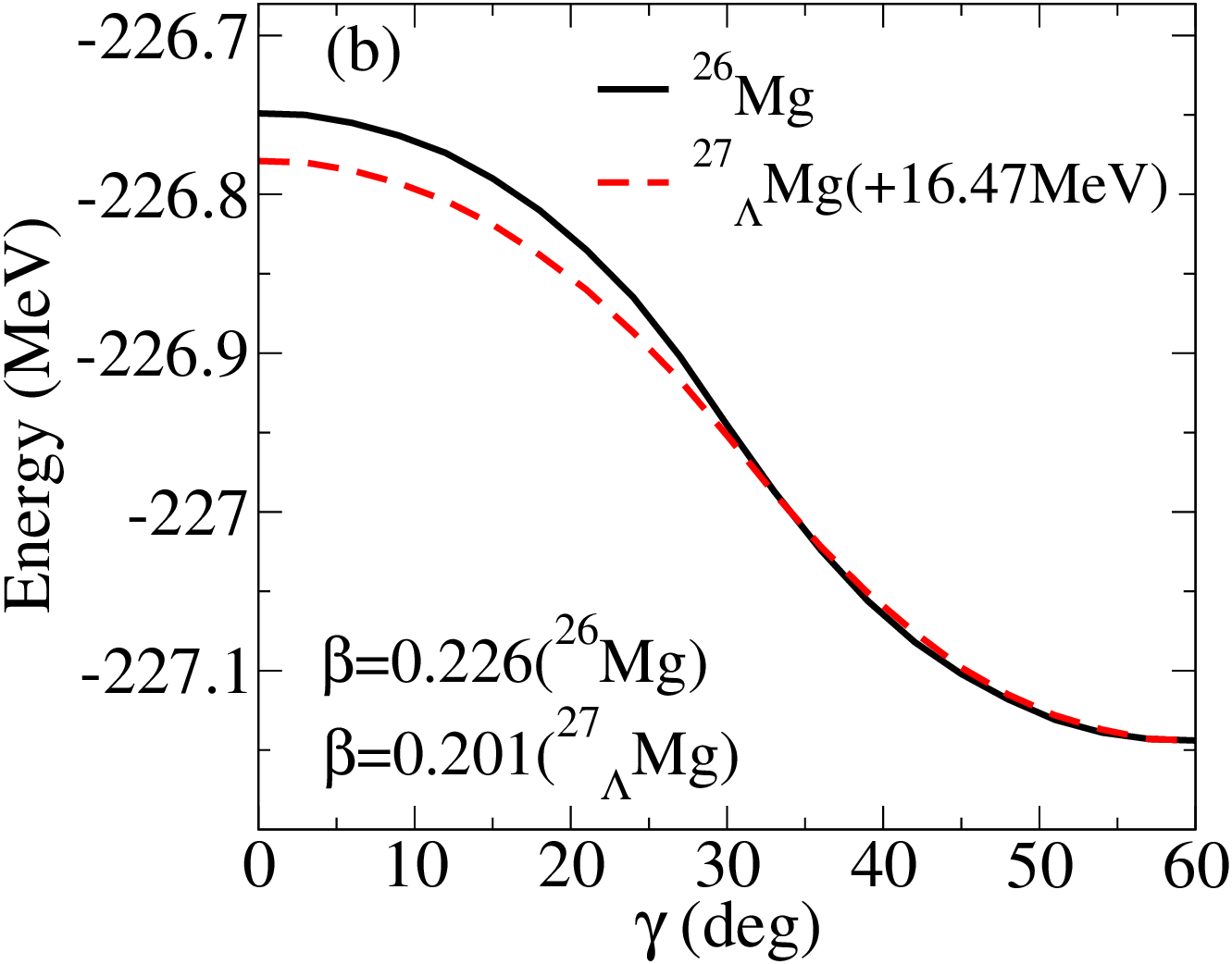}
\end{center}
\caption{(Color online) (a) The energy curve for $^{26}$Mg and 
$^{27}_{~\Lambda}$Mg 
along the axially symmetric deformation corresponding to Fig. 12. 
(b) The energy curve as a function of the 
triaxial deformation parameter $\gamma$ for the optimum values of 
the $\beta$ deformation parameters shown in Fig. 12. The energy 
surfaces for  $^{27}_{~\Lambda}$Mg are shifted by a constant amount as
indicated in the figures.}
\label{26axial}
\end{figure}

We next discuss the $^{27}_{~\Lambda}$Mg and $^{27}_{\Lambda}$Si nuclei. 
$^{26}$Mg has Z=12 and N=14, that is, the protons favour a prolate 
configuration while the neutrons favour an oblate configuration. 
As a competition of these two opposite effects, the structure of 
$^{26}$Mg may not be trivial \cite{Hinohara10}. 
$^{26}$Si is the mirror nucleus of $^{26}$Mg, and the deformation 
properties are expected to be similar to $^{26}$Mg. 

Figures ~\ref{26Si}(a) and Fig.~\ref{26Mg}(a) show the 
potential energy surfaces for 
$^{26}$Si and $^{26}$Mg, respectively. 
Indeed, the two energy surfaces are similar to each other, 
and show an oblate minimum with a considerably flat surface 
along the $\gamma$ direction. 
The energy difference between the 
oblate and the prolate configurations is 
0.12 MeV for $^{26}$Si and 0.39 MeV for 
$^{26}$Mg. 
The energy curve for $^{26}$Mg 
as a function of $\beta$ with $\gamma=0$, and 
of $\gamma$ with $\beta=\beta_{\rm min}=0.226$ 
are plotted in 
in Figs.~\ref{26axial}(a) and ~\ref{26axial}(b), respectively. 
The energy curves for $^{26}$Si are qualitatively similar, and  
are not shown. 

As in $^{24}$Mg and $^{28}$Si, 
the addition of a $\Lambda$ hyperon does not significantly alter 
the potential energy 
surface of these nuclei, 
although it somewhat softens 
the energy surface along the $\gamma$ direction (see Figs. 
~\ref{26Si}(b),~\ref{26Mg}(b), and ~\ref{26axial}). 
We again find that the additional $\Lambda$ particle favours the 
spherical configuration. 

The calculations presented in this subsection are performed 
with 
the SGII set of the Skyrme interaction. 
We have repeated the same calculations with another Skyrme parameter, SIII, 
and have found that the results are qualitatively the same as 
the results obtained with SGII. 

\subsubsection{$\gamma$ vibration}

For a long time, the $^{24}$Mg nucleus has been considered to be 
a candidate of nuclei with a triaxial shape, 
because
of the low-lying second 2$^+$state in the 
rotational spectrum \cite{Robinson68}. 
That is, the experimental spectrum of $^{24}$Mg 
has been 
interpreted as consisting of a $K$=0 
ground state rotational band and a $K$=2 rotational band built upon a $\gamma$
vibrational state at 4.23 MeV \cite{Cohen62,Batchelor60}.
The previous mean-field calculations for 
the ground state of $^{24}$Mg using RMF \cite{Koepf88} and SHF \cite{Bonche87}
have shown an axially symmetric prolate ground state. 
Recently, 
it has been pointed out that the angular momentum projection 
is essential in order to reproduce 
the triaxial ground state of 
$^{24}$Mg\cite{Yao10}. 
The 3-dimensional angular momentum projection plus generator coordinate 
method (3DAMP+GCM) calculations 
have shown a good agreement with 
the experimental data on low-spin states of $^{24}$Mg \cite{Yao10}. 

In order to see the effect of $\Lambda$ hyperon on $\gamma$ vibration, 
we
compute the second derivative of the energy curve with respect 
to $\gamma$ around the minimum, $(\beta_0,\gamma_0)$. 
That is, when we approximate the energy curve around the 
minimum as, 
\begin{equation}
E(\beta_0,\gamma)\approx 
E(\beta_0,\gamma_0)+\frac{1}{2}\,D\omega^2\,\gamma^2,
\end{equation}
the second derivative $E''$ provides information on the 
frequency $\omega$ for the $\gamma$ vibration if the vibrational 
moment of inertia $D$ is known. 
By numerically taking the second derivative, we obtain 
$E''_{^{25}_{~\Lambda} Mg}/E''_{^{24}Mg}$ = 0.961
and $E''_{^{27}_{~\Lambda} Mg}/E''_{^{26}Mg}$ = 0.988. 
That is, 
the addition of a $\Lambda$ particle makes the $\gamma$ vibration 
softer if the change in the moment of inertia $D$ is negligible. 

Although the $\Lambda$ particle softens the energy curve 
both for $^{25}_{~\Lambda}$Mg and $^{27}_{~\Lambda}$Mg, the mechanism 
is somewhat different between the two nuclei. For the $^{27}_{~\Lambda}$Mg 
nucleus, 
the prolate configuration decreases more energy as compared 
to the oblate configuration (see Fig. 13 (b)), 
because the prolate configuration has a larger overlap between 
the $\Lambda$ particle and the nucleon densities, 
as is discussed in the Appendix. 
As a consequence, the curvature of the energy curve at the 
oblate minimum becomes smaller with the 
addition of a $\Lambda$ particle. 
In contrast, the effect of the smaller value of $\beta$ is more 
significant in $^{25}_{~\Lambda}$Mg. That is, 
the energy curve along the axially symmetric configuration is 
considerably steep for $\beta \leq -0.2$ 
(see Fig. 10 (a)), and even a small change in $\beta$ induces 
a significant energy change 
at the oblate configuration. Therefore, for $^{25}_{~\Lambda}$Mg, 
the energy decreases more at the oblate side as compared to the 
prolate side for a fixed value of $\beta$ (see Fig. 10 (b)), 
leading to the softer gamma-vibration. 
Notice that the absolute value of $\beta$ is relatviely small for $^{26}$Mg, 
lying in a ``flat'' region, and this effect is much 
less important in the $^{27}_{~\Lambda}$Mg nucleus.


\begin{table*}[hbt]
\caption{
The quadrupole deformation parameters $\beta$ and $\gamma$ (in deg.), and 
the ground state energy  
obtained with the Skyrme interaction SGII parameter set. 
$\beta_p$, $\beta_n$, $\beta_{\Lambda}$, and $\beta_{\rm tot}$  
($\gamma_p$, $\gamma_n$, $\gamma_{\Lambda}$, and $\gamma_{\rm tot}$)
are the deformation parameters for proton, neutron, $\Lambda$ particle, and 
the whole nucleus, respectively. 
}
\begin{center}
\begin{tabular}{c|cccc|cccc|c}
\hline
\hline
 nucleus &  $\beta_p$  &  $\beta_n$  &  $\beta_{\Lambda}$  &  $\beta_{\rm tot}$  &  $\gamma_p$  &  $\gamma_n$  &  $\gamma_{\Lambda}$  &  $\gamma_{\rm tot}$  
&  $-E$ (MeV)  \\ 
\hline
$^{10}$C & 0.182 & 0.175 & -  & 0.35 & 0 & 0 & - & 0 & 63.222 \\
$^{11}_{~\Lambda}$C & 0.140 & 0.087 & 0.010 & 0.223 & 59.619 & 60 & 59.723 & 59.989 & 72.573  \\
\hline
$^{12}$C & 0.154 & 0.151 & -  & 0.301 & 60 & 59.969 & - & 60 & 91.6 \\
$^{13}_{~\Lambda}$C & 0.145 & 0.142 & 0.012 & 0.283 & 59.971 & 60 & 59.657 & 59.657 & 102.468 \\
\hline
$^{14}$C & 0 & 0 & -  & 0 & - & -  & - & - & 112.496 \\
$^{15}_{~\Lambda}$C & 0 & 0 & 0 & 0 & - & - & - & - & 124.962 \\
\hline
$^{16}$C & 0.084 & 0.224 & -  & 0.296 & 0 & 0 & - & 0 & 118.354 \\
$^{17}_{~\Lambda}$C & 0.063 & 0.167 & 0.005 & 0.224 & 0 & 0 & 0 & 0 & 136.242 \\
\hline
$^{18}$C & 0.090 & 0.254 & -  & 0.342 & 26.078 & 15.290 & - & 17.996 & 124.478 \\
$^{19}_{~\Lambda}$C & 0.085 & 0.243 & 0.007 & 0.326 & 25.940 & 15.368 & 19.871 & 18.045 & 138.102 \\
\hline
$^{20}$C & 0.083 & 0.234 & -  & 0.315 & 47.96 & 11.62 & - & 59.58 & 131.156 \\
$^{21}_{~\Lambda}$C & 0.083 & 0.237 & 0.007 & 0.318 & 60 & 59.840 & 60 & 60 & 145.325 \\
\hline
$^{22}$C & 0.064 & 0.185 & -  & 0.249 & 60 & 59.756 & - & 60 & 132.779 \\
$^{23}_{~\Lambda}$C & 0.061 & 0.172 & 0.005 & 0.232 & 60 & 59.770 & 60 & 60 & 147.551 \\
\hline
\hline
$^{26}$Si        & 0.1 & 0.128 & -  & 0.224 & 60 & 59.988 & 60 & - & 216.938 \\
$^{27}_{~\Lambda}$Si 
                & 0.129 & 0.1 & 0.004 & 0.224 & 60 & 59.900 & 60 & 60 & 233.373 \\
\hline
$^{28}$Si & 0.144 & 0.140 & -  & 0.278 & 59.979 & 60 & - & 60 & 245.900 \\
$^{29}_{~\Lambda}$Si & 0.132 & 0.128 & 0.004 & 0.255 & 59.980 & 60 & 60 & 60 & 262.773 \\
\hline
$^{24}$Mg & 0.2 & 0.195 & - & 0.387 & 0 & 0 & -  & 0 & 206.924 \\
$^{25}_{~\Lambda}$Mg & 0.190 & 0.185 & 0.007 & 0.368 & 0 & 0 & 0 & 0 & 222.714 \\
\hline
$^{26}$Mg & 0.104 & 0.126 & - & 0.226 &  59.951 & 60 & - & 60 & 227.144 \\
$^{27}_{~\Lambda}$Mg & 0.093 & 0.112 & 0.003 & 0.201 & 59.954 & 60 & 60 & 60 & 243.609 \\
\hline
\hline
\end{tabular}
\end{center}
\end{table*}

\section{Summary}

We have investigated the shape of $\Lambda$ hypernuclei in the 
($\beta,\gamma$) 
deformation
plane with the Skyrme-Hartree-Fock + BCS approach. 
In contrast to the previous mean-field studies, we have 
taken into account the triaxial deformation using 
a 3-D Cartesian mesh method. 
We have studied the potential energy surface 
for the Carbon hypernuclei from $^{11}_{~\Lambda}$C to 
$^{23}_{~\Lambda}$C as well as sd-shell hypernuclei 
 $^{27}_{~\Lambda}$Si, $^{29}_{~\Lambda}$Si, $^{25}_{~\Lambda}$Mg 
and $^{27}_{~\Lambda}$Mg. 
The potential energy surface for $^{10}$C, $^{26}$Mg, and $^{26}$Si 
is characterized by a flat surface along the $\gamma$ degree of freedom 
connecting the prolate and the oblate configurations. 
We have found that the addition of a $\Lambda$ particle makes the 
energy surface slightly softer along the triaxial degree of freedom, 
although the gross feature of the energy surface 
remains similar to the energy surface for the corresponding core nuclei. 

In Refs. \cite{Myaing08,Schulze10}, we have argued that the influence of the 
addition of a $\Lambda$ particle is stronger in the relativistic mean-field 
approach as compared to the non-relativistic Skyrme-Hartree-Fock approach 
employed in this paper. 
This implies that the softening of the energy curve 
for the $\gamma$-vibration 
may be larger than that estimated in this paper, if we employ 
the RMF approach instead of the SHF approach. 
It would be an interesting future subject to carry out three-dimensional calculations 
for hypernuclei with RMF in order to confirm whether it is the case. 

There would be many ways to improve our calculations presented in this paper. 
Firstly, 
as the angular momentum projection is shown to be 
essential to yield the triaxial shape of $^{24}$Mg \cite{Yao10}, 
it may be important to carry out 
the angular momentum projection on top of the mean-field energy surface 
in order to discuss the effect of $\Lambda$ particle on the shape 
of hypernuclei. 
Secondly, for nuclei with a flat energy surface along the $\gamma$ 
direction, 
the generator coordinate method may be required. In particular, it 
will provide a more quantitative estimate for the energy change 
of the $\gamma$ vibrational state due to the addition of 
$\Lambda$ particle. 
In any case, the mean-field calculations presented in this paper 
provide a good starting point for these calculations.

It will be an interesting subject to experimentally measure the 
deformation properties and collective motions of hypernuclei. 
A discussion has been started for a future experiment 
of $\gamma$-ray spectroscopy of sd-shell hypernuclei 
at the new generation 
experimental facilities, {\it e.g.,} the J-PARC facility
\cite{Koike08,Tamura09}. 
The change of deformation will be well investigated if excitation energies 
in a rotational band and the B(E2) values can be measured experimentally.

\begin{acknowledgments}
We thank H. Tamura, Y. Zhang, N. Hinohara and A. Ono for 
useful discussions. 
This work was supported by the Japanese
Ministry of Education, Culture, Sports, Science and Technology
by Grant-in-Aid for Scientific Research under
the program number 22540262.
\end{acknowledgments}

\appendix

\section{Overlap between $\Lambda$-particle and nucleon densities} 

\begin{figure}[htb]
\includegraphics[clip,scale=0.5]{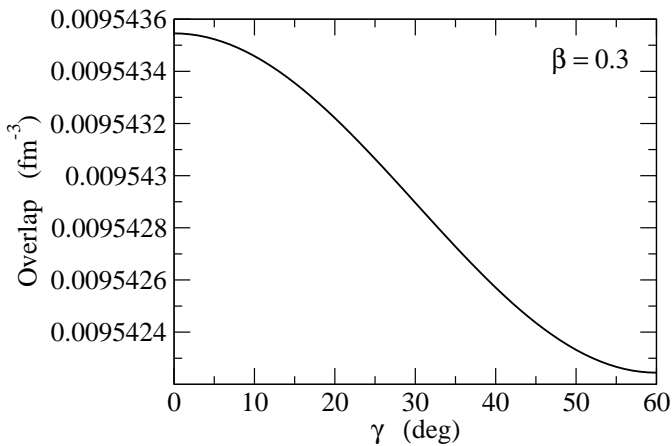}
\caption{
The overlap between the nucleon and the $\Lambda$ particle densites 
as a function of the triaxiality $\gamma$ for a fixed value of $\beta=0.3$. 
The $\Lambda$-particle density is assumed to be a spherical Gaussian function, while 
a deformed Woods-Saxon shape is considered for the nucleon density. }
\end{figure}

In this Appendix, we discuss the overlap between $\Lambda$ particle and nucleon density distributions 
using simple parametrizations for the density distributions. 
Since we consider that a $\Lambda$ particle occupies the lowest single-particle state, we 
assume that the $\Lambda$ particle density is almost spherical and is given by, 
\begin{equation}
\rho_\Lambda(r)=\frac{1}{(\sqrt{\pi}b)^3}\,e^{-r^2/b^2}. 
\end{equation}
On the other hand, for the nucleon density, we assume that it is given by a deformed 
Woods-Saxon form, that is, 
\begin{equation}
\rho_N(r,\theta,\phi)=\frac{\rho_0}{1+\exp[(r-R(\theta,\phi))/a]},
\end{equation}
where
\begin{eqnarray}
R(\theta,\phi)&=&R_0(\beta,\gamma)\cdot[1+\beta\cos\gamma\,Y_{20}(\theta) \nonumber \\
&& +\frac{1}{\sqrt{2}}\beta\sin\gamma\,(Y_{22}(\theta,\phi)+Y_{2-2}(\theta,\phi))].
\end{eqnarray}
Here, the radius 
$R_0(\beta,\gamma)$ is determined for each $\beta$ and $\gamma$ in order to 
satifsy the volume conservation condition, that is, 
\begin{equation}
F_{\rm vol}=\int^\infty_0r^2dr\int^1_{-1}d(\cos\theta)\int^{2\pi}_0d\phi\,\frac{1}{1+\exp([r-R(\theta,\phi))/a]},
\end{equation}
is independent of $\beta$ and $\gamma$. 

Figure 14 shows the overlap between $\rho_\Lambda$ and $\rho_N$, that is, 
\begin{equation}
O=\int^\infty_0r^2dr\int^1_{-1}d(\cos\theta)\int^{2\pi}_0d\phi\,\rho_\Lambda(r)\rho_N(r,\theta,\phi),
\end{equation}
as a function of the triaxiality $\gamma$ 
for a fixed value of $\beta=0.3$. To this end, we use $R_0(\beta=0,\gamma=0)=1.1\times 24^{1/3}$ fm, 
and $a$=0.55 fm. The value of $\rho_0$ is fixed so that the volume integral of $\rho_N$ is 24. 
We use $b$=1.565 fm, which corresponds to the harmonic oscillator with a frequency of 
$\hbar\omega=41\times 24^{-1/3}$ MeV. 
As one can see, the overlap is the largest for the prolate configuration, although the variation is small 
with respect to $\gamma$.


\end{document}